  \documentclass[traditabstract]{aa}  

\usepackage{graphicx}
\usepackage{natbib}
\usepackage{lscape}
\usepackage{supertabular}
\usepackage{xcolor}

\newcommand{\pa}[1]{\mbox{Pa\,#1}}
\newcommand{\pab}[0]{\mbox{Pa$\beta$}}

\usepackage{txfonts}
\begin{document}
\title{The CARMENES search for exoplanets around M dwarfs}
\subtitle{Behaviour of the Paschen lines during flares and quiescence\thanks{Full Table 2 is only available in 
electronic form
at the CDS via anonymous ftp to cdsarc.u-strasbg.fr (130.79.128.5)
or via http://cdsweb.u-strasbg.fr/cgi-bin/qcat?J/A+A/}}

\author{B. Fuhrmeister\inst{1}, S. Czesla\inst{\ref{inst13}}  
  \and J. H. M. M. Schmitt\inst{\ref{inst1}}
    \and  P.~C. Schneider\inst{\ref{inst1}}
    \and  J.~A.~Caballero\inst{\ref{inst3}}
  \and  S.~V.~Jeffers\inst{\ref{inst15}}
  \and  E.~Nagel\inst{\ref{inst2}}
  \and  D.~Montes\inst{\ref{inst10}}
  \and  M.~C.~G\'alvez~Ortiz \inst{\ref{inst3}}
  \and   A.~Reiners\inst{\ref{inst2}}
  \and   I.~Ribas\inst{\ref{inst5},\ref{inst4}}
  \and  A.~Quirrenbach\inst{\ref{inst7}}
  \and P.~J.~Amado\inst{\ref{inst6}}
  \and Th.~Henning\inst{\ref{inst14}}
  \and  N.~Lodieu\inst{\ref{inst8},\ref{inst9}} 
  \and  P.~Mart\'in-Fern\'andez\inst{\ref{inst12}}
  \and  J.~C.~Morales\inst{\ref{inst5},\ref{inst4}}
  \and  P.~Sch\"ofer\inst{\ref{inst6}}
  \and  W.~Seifert\inst{\ref{inst7}} 
  \and M.~Zechmeister\inst{\ref{inst2}}
  }

\institute{Hamburger Sternwarte, Universit\"at Hamburg, Gojenbergsweg 112, 21029 Hamburg, Germany\\
  \email{bfuhrmeister@hs.uni-hamburg.de}\label{inst1}
       \and
     Th\"uringer Landessternwarte Tautenburg, Sternwarte 5, 07778 Tautenburg, Germany\label{inst13} 
        \and
        Centro de Astrobiolog\'{\i}a, CSIC-INTA, Camino Bajo del Castillo s/n, 28692 Villanueva de la Ca\~nada, Madrid, Spain \label{inst3}
     \and
        Max-Planck-Institut f\"ur Sonnensystemforschung, Justus-von-Liebig-Weg 3,37077 G\"ottingen, Gemany\label{inst15}
        \and
        Institut f\"ur Astrophysik, Friedrich-Hund-Platz 1, 37077 G\"ottingen, Germany\label{inst2} 
        \and
        Facultad de Ciencias F\'{\i}sicas, Departamento de F\'{\i}sica de la Tierra y Astrof\'{\i}sica; IPARCOS-UCM (Instituto de F\'{\i}sica de Part\'{\i}culas y del Cosmos de la UCM), Universidad Complutense de Madrid, 28040 Madrid, Spain\label{inst10} 
        \and
       Institut d'Estudis Espacials de Catalunya, 08034 Barcelona, Spain\label{inst5}
        \and  
        Institut de Ci\`encies de l'Espai (CSIC), Campus UAB, c/ de Can Magrans s/n, 08193 Bellaterra, Barcelona, Spain\label{inst4}
        \and
        Landessternwarte, Zentrum f\"ur Astronomie der Universit\"at Heidelberg, K\"onigstuhl 12, 69117 Heidelberg, Germany\label{inst7} 
        \and
        Instituto de Astrof\'isica de Andaluc\'ia (CSIC), Glorieta de la Astronom\'ia s/n, 18008 Granada, Spain\label{inst6} 
        \and
        Max-Planck-Institut f\"ur Astronomie, K\"onigstuhl 17, 69117 Heidelberg, Germany\label{inst14}
        \and
        Instituto de Astrof\'{\i}sica de Canarias, c/ V\'{\i}a L\'actea s/n, 38205 La Laguna, Tenerife, Spain\label{inst8}
        \and
	Departamento de Astrof\'{\i}sica, Universidad de La Laguna (ULL), 38206 La Laguna, Tenerife, Spain\label{inst9} 
        \and
        Centro Astron\'omico Hispano en Andaluc\'ia, Observatorio Astron\'omico de Calar Alto, Sierra de los Filabres, 04550 G\'ergal, Almer\'{\i}a, Spain\label{inst12} 
}

\date{Received 12/06/2023; accepted 07/08/2023}

\abstract
    {The hydrogen Paschen lines are known activity indicators, but  
    studies of them in M~dwarfs during quiescence are as rare as their 
    reports in flare studies. 
    This situation is mostly caused by a lack of observations, owing to their location in the near-infrared regime, which is covered by few high-resolution spectrographs.
    We study the Pa$\beta$ line,
    using a sample of 360 M~dwarfs observed by the CARMENES spectrograph.
    Descending the spectral sequence of inactive M~stars in quiescence, we find the \pab\ line to get shallower until
    about spectral type M3.5\,V, after which a slight re-deepening is observed. 
    Looking at the whole sample, for stars with H$\alpha$ in absorption,
    we find a loose anti-correlation between the (median) pseudo-equivalent widths (pEWs) of H$\alpha$ and 
    Pa$\beta$ for stars of similar effective temperature.
    Looking instead at time series of 
    individual stars, we often find correlation between pEW(H$\alpha$) and pEW(Pa$\beta$)
    for stars with H$\alpha$ in emission and an anti-correlation for stars with
    H$\alpha$ in absorption.
    Regarding flaring activity, we report the
    automatic detection of 35 Paschen line flares in 20 stars. Additionally we found
    visually six faint Paschen line flares in these stars plus 16 faint Paschen line flares
    in another 12 stars.  In strong flares, Paschen lines can be observed up to
    Pa\,14. Moreover, we find that Paschen line emission is almost always coupled to
    symmetric H$\alpha$ line broadening, which we ascribe to Stark broadening, indicating
    high pressure in the chromosphere. Finally we report a few \pab\ line asymmetries for
    flares that also exhibit strong H$\alpha$ line asymmetries.}
\titlerunning{Paschen lines in flares and quiescence}
\authorrunning{B. Fuhrmeister et~al.}
\maketitle


\section{Introduction}

Stellar activity is ubiquitous in M~dwarfs. It is frequently studied
by observing activity tracers, such as the X-ray flux \citep{Pizzolato2003, Foster2022, Magaudda2022} 
or chromospheric line fluxes \citep{GomesdaSilva2021}. 
Many lines sensitive to the chromosphere and transition region are 
located in the ultraviolet (UV), such as the Ly$\alpha$ line \citep{MUSCLES2016},
which, together with the bulk UV emission, is a crucial ingredient to assess
the habitability of exoplanets \citep{MUSCLES2017}. However, all important
UV lines are not observable
from the ground and partly not even with current satellites. 
Fortunately, also the observable visible range includes a number of activity-sensitive lines,
such as the \ion{Ca}{ii} H\&K lines and the related near-infrared (NIR) triplet lines of \ion{Ca}{ii},
which are not as sensitive to activity as the former though
\citep{Martinez-Arnaiz2011,Johannes,Mittag2017,Pavlenko2019}.
Yet, the most widely used activity indicator in M~dwarfs
remains the H$\alpha$ line \citep{Gizis2002,Walkowicz2009,Lodieu2011,Patrick}.

Extensive studies of spectral activity tracers in the infrared  have only recently
become feasible with the advent of NIR high-resolution spectrographs, such as
CARMENES \citep{Quirrenbach2020}. Although chromospheric indicator lines
are often less prominent in the infrared, there are a few known examples such as
the \ion{K}{i} doublet, which was reported to be magnetically
sensitive \citep{kalium, Terrien2022}, and the \ion{He}{i} infrared triplet (IRT), which is a long-known activity
tracer in the Sun and other stars \citep{Zirin1968,Sanz-Forcada2008,Andretta2017}. 
The Paschen (Pa) series of hydrogen, which
is known to react to strong flares, is also located in the infrared regime. 
In the solar context, Paschen lines were used mainly to determine the
electric field in prominences via the Stark effect \citep{Foukal1987,Casini1996}.
For stars,
there are a number of flare observations, in which members of the Pa
series were detected in emission. For example,
\citet{Liebert1999} observed the M9.5 dwarf 2MASSW~J0149090+295613 during a mega-flare event,
which allowed them to detect \pa{7}\footnote{We use the notation \pa{$N$} to indicate the Paschen line with
the upper level $N$ and only refer to \pab, Pa$\gamma$, and Pa$\delta$, 
which are equivalent to \pa{5}, Pa\,6 and Pa\,7, using the Greek nomenclature, unless in direct comparison to higher level lines.}
 through 11 in emission, and observe their decay
during another three observations within about 30~min. 
\citet{Schmidt2007} observed a large flare on the M7 star 2MASS~J1028404$-$143843,
which showed strong continuum enhancement even at 10\,000~\AA\ and exhibited Pa\,8 to 11
in emission.
\citet{CNLeoflare} found Pa\,7 to 11 in emission during a mega-flare on CN~Leo (M5.5),
which decayed fully within about 10~minutes.
The \pa{7} to 9 lines were also reported in emission for a medium sized flare on Proxima Centrauri by \citet{proxcen}.
In a dedicated flare search covering about 50~hours of observation time of the active M~dwarfs EV~Lac, AD~Leo, YZ~CMi, and
vB~8, \citet{Schmidt2012} detected 16 flares, out of which three showed
infrared emission including Pa\,5 (= Pa$\beta$) and Pa\,6 (= Pa$\gamma$). 
\citet{Kanodia2022} analysed high-resolution flare spectra of the M8 dwarf
vB~10 covering \pa{6}, 7, 11, and 12. While \pa{12} was only marginally detected, 
\pa{6} showed indications of a weak red asymmetry and persisted to be in emission 
in a second exposure about 20 minutes later.
In a study of H$\alpha$ line asymmetries using CARMENES data, \citet{asym} searched for \pab\
line emission in 36 spectra exhibiting flare-induced H$\alpha$ wing asymmetries and reported
9 weak detections of \pab\ emission.

Examples of non-flare studies are more rarely found.
In a publication by \citet{Klein2020}, who studied the M1 dwarf AU~Mic, it was found that the
stellar rotation period could be recovered using the \ion{He}{i} IRT and the
\pab\ lines. These authors also reported that the origin of the \ion{He}{i} IRT lines seems to be
more concentrated toward equatorial latitudes, while the \pab\ line is primarily formed at polar regions on AU Mic.
Turning to large stellar samples, \citet{Patrick} used CARMENES data in a comprehensive
activity study to compare the relation between different activity indicators
in M~dwarfs using a spectral subtraction technique. While they did not find any correlation between
the pseudo equivalent-width (pEW) of \pab\
and that of the H$\alpha$ line, they reported deeper (excess) absorption in the \pab\ line
for the most active stars as measured by H$\alpha$ and of spectral type earlier than M4.0.

In the following, we utilize the unique database of M~dwarf spectra obtained with the CARMENES spectrograph
to study the Paschen lines along the M~dwarf sequence in more
detail. First, we analyse their behaviour
in the quiescent activity levels of the stars along the M sequence 
and, second, we investigate the Paschen lines for the detected flares.
Specifically, we address the question of how often flares are detectable by Paschen emission in
comparison to H$\alpha$ emission and what physical parameters favour Paschen line emission.


\section{Observations}\label{sec:obs}

All spectra used in this study were taken
with the CARMENES spectrograph, installed at the 3.5\,m Calar Alto 
telescope \citep{Quirrenbach2020}.
CARMENES covers the wavelength range
from 5\,200 to 9\,600\,\AA\, in the visual channel (VIS) and from 9\,600 to 17\,100\,\AA\, in
the near-infrared
channel (NIR). The instrument provides a spectral resolution of
$\sim$ 94\,600 in VIS and $\sim$ 80\,400 in NIR. 
While the CARMENES data are obtained mainly for planet search,
they are also a resource for studies of stellar
parameter determination and activity. A large part of the data (years 2016--2020) have
become public \citep{carmenes2023}.
The data are especially well suited also for other purposes, since the CARMENES
sample is biased only marginally. Since {\it Gaia} data were not available at the time
of building the CARMENES guranteed time observations M-dwarf sample
\citep{Alonso2015,Reiners2017,carmenes2023}, the CARMENES consortium  selected the brightest
stars (in $J$ band) for each spectral subtype that were observable from Calar Alto (i.\,e.
$\delta >$ --23\,deg) and that did not have any known close companion at
$\rho <$ 5\,arcsec. As a result, the only bias in our sample is Malmquist's, by which
overluminous young stars in stellar kinematic groups are over-represented in
our target list. However, most of these are  very active and have, therefore, a large RV
jitter that impedes reaching the main scientific objective of
CARMENES, which is the search for Earth-like planets in the habitable zone of M
dwarfs.
As a result, the consortium discontinued observations of a few ``RV-loud'' stars at the
beginning of the program, after a minimum number observations \citep{Lev}.
Nevertheless, we also include 27 of the 31 ``RV-loud'' stars in the investigation in this work.

In our analysis, we considered a
sample of 360 M~dwarfs observed by CARMENES, resulting in
more than 19\,000 spectra taken before September 2022. We excluded known binaries
\citep{Baroch2018,Schweitzer2019,Baroch2021}, which may
hamper our analysis by orbit-induced line shifts.
Moreover, the CARMENES consortium has invested a considerable effort in determining
stellar parameters of the target stars, like spectral
types \citep{Alonso2015}, luminosities and colours \citep{Cifuentes2020}, photospheric
parameters, namely $T_{\rm eff}$, $\log{g}$, and [Fe/H]
\citep{Passegger2018,Passegger2019,Marfil2021,Passegger2022}, 
rotation velocities \citep{Reiners2017}, rotation
periods \citep{DA19,ytong}, magnetic fields
\citep{Reiners2022}, and masses and radii \citep{Schweitzer2019}. Furthermore,
stellar activity has been studied using
the CARMENES high resolution spectroscopic data. In particular, H$\alpha$ was
investigated \citep{asym,Patrick}, along with other activity
sensitive lines. Examples are the \ion{He}{i} infrared triplet \citep{hepaper,hevar} or
the optical and infrared \ion{K}{i} doublets \citep{kalium}. Also other spectral indicators 
\citep{Zechmeister2018,Patrick,Schoefer2022} were studied. In this work we make extensive
use of the results obtained in these publications and refer to them for further details.

The stellar spectra
were reduced using the CARMENES reduction pipeline
\citep{pipeline,Caballero2}. Subsequently, we corrected them for barycentric and
systemic radial velocity motions and carried out a correction for telluric absorption lines \citep{Evangelos}
using the {\tt molecfit}
package\footnote{\tt{https://www.eso.org/sci/software/pipelines/skytools /molecfit}}.
No correction for airglow emission lines was attempted, although they can play a role
near the Paschen lines and may be shifted into the integration ranges used for their
analysis. Because of the amount of available data and the difficulty in dealing with the
lines automatically, we decided
to identify and remove the affected spectra later on in the analysis.

\begin{figure}
\begin{center}
\includegraphics[width=0.5\textwidth, clip]{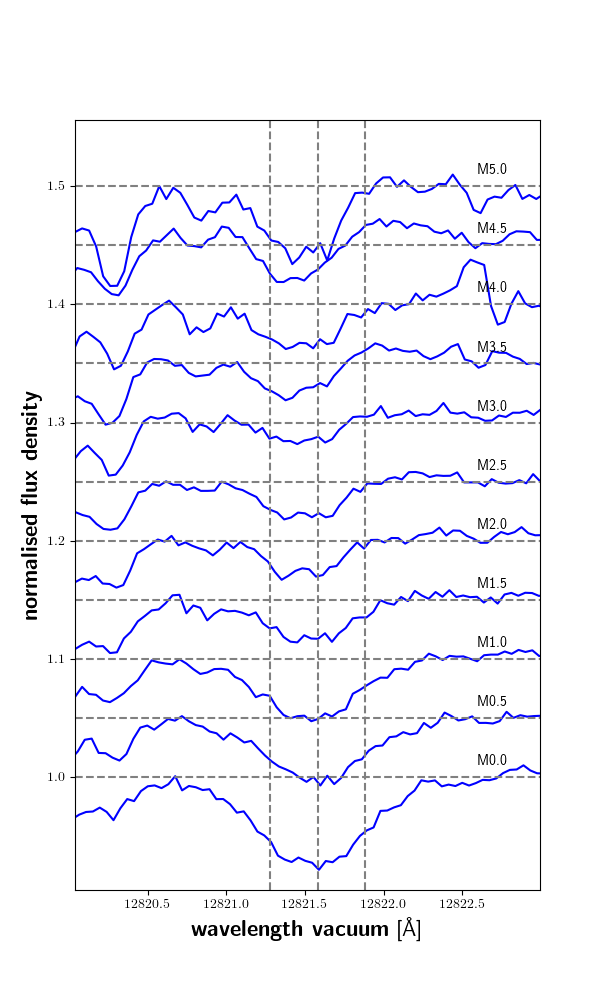}\\
\caption{\label{Mseq} Spectral subtype sequence of the wavelength region around the Pa$\beta$ line
  for stars with H$\alpha$ in absorption. Spectra of stars later than M5.0\,V are
  not shown, since they either show H$\alpha$ emission or have a low signal-to-noise.
  Each normalised spectrum is offset for convenience (offset marked as a horizontal dashed line).
  The dashed vertical lines mark the central wavelength
  of the Pa$\beta$ line and the lower and upper limit of the line integration band. From bottom to top the following stars are shown:
  J03463+262 / HD\,23453 (M0.0\,V), J02222+478 / BD+47\,612 (M0.5\,V), J00051+457 / GJ\,2 (M1.0\,V), J13196+333 / Ross\,1007 (M1.5\,V), J01013+613 / GJ\,47 (M2.0\,V),
  J00389+306 / Wolf\,1056 (M2.5\,V), J02015+637 / G\,244-047 (M3.0\,V), J12479+097 / Wolf\,437 (M3.5\,V), J04311+589 / STN\,2051A (M4.0\,V), J08119+087 / Ross\,619 (M4.5\,V)
  J18165+048 / G\,140-051 (M5.0\,V).
}
\end{center}
\end{figure}

The CARMENES instrument does not cover the Pa\,$\alpha$ line at 18756.4 \AA, but all
other members of the Paschen series are covered, including Pa\,5 (\pab) at 12\,821.578\,\AA,
\pa{6} (Pa\,$\gamma$) at 10\,941.17\,\AA, and \pa{7} (Pa\,$\delta$) at 10\,052.6\,\AA. The Paschen series ends
at about 8250\,\AA, which is located already in the VIS channel of CARMENES.
O$_{2}$ airglow lines are found near the \pab\ line at (vacuum) wavelengths of 12\,819.46, 12\,822.43, 12\,824.78~\AA, with the
latter line being the strongest \citep{Oliva2015}.

\section{Analysis of activity indicators and flare detection}

\subsection{pEW measurements of activity indicators}\label{sec:pewmeasure}

To assess the activity state of the stars in each spectrum, we employed pEW measurements.
The spectra of M dwarfs do not show an identifiable continuum because of the abundance of molecular
absorption lines. These pEW measurements were then used to search for flares in H$\alpha$ and \pab.

To give an overview of the appearance of the \pab\ line, we show examples of it along
the M dwarf sequence in Fig. \ref{Mseq}
for stars with H$\alpha$ in absorption. In this sequence, the strongest \pab\ absorption is observed in
the M0.0 star. Absorption subsequently weakens until a minimum is reached for the shown M3.5 star, after which
the \pab\ lines deepens again; a more detailed discussion is provided in Sect.~\ref{sec:inactive}.

To quantify the level of absorption or emission in the lines,
we computed pEWs of the H$\alpha$ line, the bluest and middle lines of the \ion{Ca}{ii} IRT,
as well as the Pa$\beta$, Pa$\gamma$, and Pa$\delta$ lines. We considered H$\alpha$
to be in absorption if the pEW value was larger than $-0.6$\,\AA, while lower values
marked H$\alpha$ in emission. This threshold was already used by \citet{hepaper} and
is in between other adopted values as $-0.5$\,\AA\ \citep{Jeffers2018} 
or $-0.75$\,\AA\ \citep{West2011}. If H$\alpha$ is in absorption or emission has
traditionally been used to discriminate between inactive or active stars and we also
use it here to split up our sample in this sense. 

For the pEW computation, we list the central wavelength, full width of the line integration window, and the location of the two
reference bands in Table~\ref{tab:ew}; for a more detailed description
of pEW measurements of chromospheric lines we refer to \citet{cycle}.
While the reference bands are typically blue- and
redward of the central wavelength, both are located blueward for the Pa$\beta$ line,
since it is located near the red edge of one of the spectral CARMENES orders and we did not 
want to use a reference band in a different order.
While we used $1.5$~\AA\, wide line integration bands for the Pa$\gamma$ and Pa$\delta$ lines,  
we opted for
a narrower 0.6~\AA\, wide integration band of Pa$\beta$ to minimize contamination by airglow.
This is similar to the even narrower 0.5~\AA\, band used by 
\citet{Patrick} for \pab. The
widths of 1.6 \AA\, and 0.5 \AA\,
for the H$\alpha$ and \ion{Ca}{ii} IRT lines were also used by
\citet{hevar}.

For all studied lines,
emission during flares may be broader than the line integration band. Therefore,
in extreme cases, the full variability range may not be represented with our choice of integration ranges.
Moreover, rotation rates higher than about 
$v\, \sin i = 15$ km\,s$^{-1}$ will affect the pEW measurements by shifting flux out of the integration bands (this threshold is exceeded by 20 of our sample stars).
Nevertheless, we found
the chosen integration bands be suitable for identifying variability, which we 
are most interested in.

\begin{table}
        \caption{\label{tab:ew} Parameters (vacuum wavelength) of the pEW  calculation. }
\scriptsize
\begin{tabular}[h!]{lcccccccccc}
\hline
\hline
\noalign{\smallskip}

Line           & Wave-   & Width  & Reference  & Reference  \\
& length      &       & band 1 &  band 2 \\
& [\AA] & [\AA] & [\AA] & [\AA]\\
\noalign{\smallskip}
\hline
\noalign{\smallskip}
H$\alpha$         & 6564.60 & 1.6 & 6537.4--6547.9 & 6577.9--6586.4 \\
\ion{Ca}{ii} IRT$_{1}$ & 8500.35 & 0.5 & 8476.3--8486.3 & 8552.4--8554.4\\
\ion{Ca}{ii} IRT$_{2}$ & 8544.44 & 0.5 & 8576.3--8486.3 & 8552.4--8554.4\\
Pa$\beta$ (Pa 5) & 12821.58 & 0.6 & 12812.0--12814.0 & 12789.0--12792.0 \\
Pa$\gamma$ (Pa 6)      & 10941.17 & 1.5 & 10902.0--10904.0 & 10964.7--10966.7\\
Pa$\delta$ (Pa 7)      & 10052.60 & 1.5 & 10045.0--10047.0 & 10076.0--10078.0 \\

\noalign{\smallskip}
\hline

\end{tabular}
\normalsize
\end{table}

\begin{sidewaystable*}
\caption{\label{tab:allpews} Measured pEWs, their MADs, and stellar parameters.$^{a}$ }
\footnotesize
\begin{tabular}[h!]{llccccccccccccc}
\hline
\hline
\noalign{\smallskip}

	Karmn           & Name &SpT  & $T_{\mathrm{eff}}$ & log\,g & pEW(H$\alpha$)    & pEW(\pab) &  pEW(Pa\,$\gamma$)  & pEW(Pa\,$\delta$) & MAD(H$\alpha$) & MAD(\pab) & MAD(Pa\,$\gamma$) & MAD(Pa\,$\delta$) & $P_{\mathrm{rot}}$ & $v\,\sin(i)$  \\
		&        &     & [K] & &[\AA]         & [\AA]   & [\AA]   & [\AA]    & [\AA] & [\AA]  &[\AA]  & [\AA] & [day] & [km\,s$^{-1}$]\\
\noalign{\smallskip}
\hline
\noalign{\smallskip}

	J00051+457 &   BD+44 4548   &                     1.0 &   3773.0 &     5.07 &    0.350     &  0.024   &    0.000  &     0.065 &      0.016 &      0.002  &     0.000&       0.002  &       15.37 &          2.0\\
	J00067-075  &  GJ 1002                   &        5.5  &  3169.0   &   5.20  &  -0.043   &    0.032&       0.014    &   0.265    &   0.061    &   0.002  &     0.005     &  0.004      &    0.00     &      2.0\\
	J00162+198E &  LP 404-062   &                     4.0 &   3329.0    &  4.93  &   0.139   &    0.010   &    0.081  &     0.119  &     0.013   &    0.003   &    0.004 &      0.004     &   105.00     &      2.0\\
	J00183+440 &   GX And                    &        1.0 &   3603.0   &   4.99  &   0.318    &   0.007    &   0.032   &    0.059    &   0.006   &    0.001 &      0.008    &   0.002      &   45.00      &     2.0\\
	J00184+440 &   GQ And  &                          3.5 &   3318.0   &   5.20  &   0.160    &   0.014 &      0.013  &     0.137   &    0.010   &    0.001  &     0.003   &    0.002   &       0.00      &     2.0\\
	J00286-066  &  GJ 1012                   &        4.0  &  3419.0  &    4.81 &    0.168   &    0.007     &  0.040    &   0.094    &   0.008   &    0.002   &    0.003    &   0.002  &        0.00      &     2.0\\
	J00389+306 &   Wolf 1056       &                  2.5 &   3551.0   &   4.90 &    0.287 &      0.008&       0.025  &     0.076   &    0.013    &   0.003    &   0.004 &      0.002   &      50.20   &        2.0\\
	J00570+450  &  G 172-030    &                     3.0&    3488.0  &    5.04  &   0.166    &   0.012   &    0.015   &    0.105    &   0.024    &   0.002  &     0.037   &    0.003 &         0.00 &          2.0\\
	J01013+613  &  GJ 47                     &        2.0 &   3564.0 &     5.05  &   0.250   &    0.013     &  0.026    &   0.081    &   0.036    &   0.002     &  0.007     &  0.001   &      34.70    &       2.0\\
	J01019+541  &  G 218-020    &                     5.0 &   3070.0   &   5.12 &   -4.200  &     0.007  &     0.018   &    0.242   &    0.451    &   0.003    &   0.004 &      0.004 &         0.14  &        30.6\\

\noalign{\smallskip}                                                 
\hline                                             
                                                   
\end{tabular}                                      
                               
\normalsize                    

$^{a}$ The full table is provided at CDS. We show here the first ten rows as a guidance.
\end{sidewaystable*}

\subsection{Search for Paschen and H$\alpha$ line flares}\label{sec:flaresearch}

To study the Paschen lines during flares, flares with a
reaction of the Paschen lines need to be identified in the first place.
The CARMENES observing schedule does rarely produce consecutive spectra of the same star,
but observations of the same star are typically separated by some days.

To facilitate a flare search, we computed for each star the
median, $\mu$, of the pEW measurements for each chromospheric line and the median
average deviation about the median (MAD). We list these values together with some basic stellar parameters in Table~\ref{tab:allpews} for each star. The MAD yields a robust
estimator of the standard deviation. If MAD(pEW(H$\alpha$)) and $\sigma(\mbox{pEW(H$\alpha$)})$
denote the MAD and standard deviation of the time series of pEW measurements of H$\alpha$
\begin{align}
	\sigma(\mbox{pEW(H$\alpha$)}) = 1.4826 \times \mbox{MAD(pEW(H$\alpha$))} \; .\end{align}
The same nomenclature is used for the other lines.

In a first step, we searched for flares indicated by H$\alpha$ and the \ion{Ca}{ii}~IRT lines.
We accepted a spectrum as flaring (and call this an H$\alpha$ flare) in case of combined
H$\alpha$ and \ion{Ca}{ii}~IRT excursions, viz., 
\begin{align}
	\mbox{(i)} & \; \mbox{pEW(H$\alpha$)} && < \mu(\mbox{pEW(H$\alpha$)}) - 3\sigma(\mbox{pEW(H$\alpha$)}) \;\;\; \mbox{and} \nonumber \\
	\mbox{(ii)} & \;	\mbox{pEW(Ca~IRT$_{1,2}$)} && < \mu(\mbox{pEW(Ca~IRT$_{1,2}$)}) \, - \nonumber \\
	& &&  \phantom{<<} 3\sigma(\mbox{pEW(Ca~IRT$_{1,2}$)}) \; , \nonumber
\end{align}
where the condition (ii) must apply to at least one of the considered \ion{Ca}{ii} IRT lines.
Using only the H$\alpha$ line in the search worked well for inactive stars, which exhibit pronounced but
seldom flares and show a rather stable H$\alpha$ absorption line otherwise, but 
not for stars showing persistent, strong variability in H$\alpha$. Since for many of these stars,
the \ion{Ca}{ii} IRT lines showed less pronounced variations
outside of flares, we coupled the search for flares showing up in H$\alpha$
to \ion{Ca}{ii} IRT. 

Flares also showing up in the Pa$\beta$ line were identified by requiring that
condition (i) and (ii) are met, (i.\,e. the star is flaring in H$\alpha$) 
and an equivalent $3\sigma$ condition for
the pEW of the Pa$\beta$ line is fulfilled. In these so identified Pa$\beta$ flares
we additionally searched for Pa$\gamma$ and Pa$\delta$ flares, applying the
$3\sigma$ condition one more time.

By this method, most cases of spurious flares induced by statistical noise are suppressed. Additionally, the coupling with
an H$\alpha$ criterion removes spurious flares caused by airglow contamination in Pa$\beta$. 
We nevertheless inspected all Pa$\beta$ flare detections by eye. In the follwoing we call all
\pab\ flares fulfilling our flare criteria 'automatically detected' and flares that pass the visual inspection 'visually confirmed'.

\section{Results and discussion}

\subsection{The Paschen lines during quiescence}\label{sec:inactive}

\subsubsection{Origin of the Paschen lines}

In Fig.~\ref{Mseq}, we show examples of the Pa$\beta$ line for inactive stars along the
M~dwarf spectral sub-type sequence. The line is purely chromospheric in origin as can
be seen from a comparison to PHOENIX photospheric models \citep{Hauschildt1999, Husser2013, Schweitzer2019},
which we show in Fig.~\ref{fig:phoenix}. The line is absent in photospheric spectra.
Moreover, it can be seen in Fig.~\ref{Mseq} that around spectral type M2.0\,V
 another absorption feature blueward of the \pab\ line at about 12821.4\,\AA\,
 starts to emerge, deepening for later spectral types. The feature can also be seen
 in the PHOENIX photospheric spectrum for the M5.0\,V star shown in Fig.~\ref{fig:phoenix}, although it is
 not as deep as observed, which is not uncommon for a molecular feature. 

 Concerning the origin of the Paschen lines, we distinguish between the
 lines observed in absorption or in emission.
\citet{Cram1979} found that for the H$\alpha$ line the line source function
is controlled by photoionization (and recombination) for
cases where it is in absorption. Therefore, the $n=3$ level as groundstate of
the \pab\ line is populated by H$\alpha$ absorption and by recombination.
On the other hand, for flaring states, where we found the line
in emission, it must be collisionally controlled, as \citet{Cram1979} argued
for H$\alpha$ as well.

\subsubsection{The Paschen lines along the M dwarf spectral sequence}

\begin{figure}
\begin{center}
\includegraphics[width=0.5\textwidth, clip]{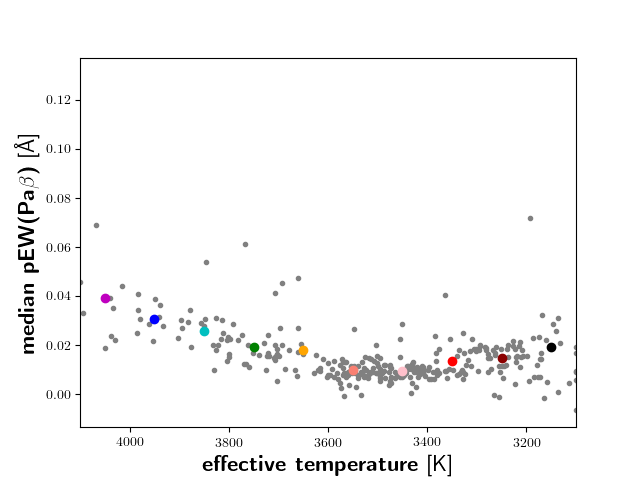}\\
\caption{\label{fig:teffpabeta}  pEW(Pa$\beta$) shown as a function of $T_{\rm eff}$.
  Grey dots represent median(pEW(Pa$\beta$)) for each star, the coloured circles
  represent the mean of these median(pEW(Pa$\beta$)) measurements for each $T_{\rm eff}$
  interval as introduced in the main text. The colours are chosen as in Figs. \ref{fig:inactive} and \ref{fig:correlation}
  to simplify comparison.
}
\end{center}
\end{figure}

To generalise the impression from the example sequence in Fig.~\ref{Mseq}, we show
in Fig.~\ref{fig:teffpabeta} the
distribution of all median(pEW(\pab)) per star as a function of the effective
temperature $T_{\rm eff}$, adopted from \citet[from spectral energy distribution fitting]{Cifuentes2020} and \citet[from spectral synthesis]{ Marfil2021}. Fig.~\ref{fig:teffpabeta}
shows that the Pa$\beta$ line becomes shallower (i.e., yields lower pEWs)
for lower $T_{\rm eff}$ (or later spectral
type) until a turning point is reached at about $T_{\rm eff} < 3400$\,K, which corresponds to spectral types of
about M4.0\,V. 

\begin{figure*}
\begin{center}
\includegraphics[width=0.5\textwidth, clip]{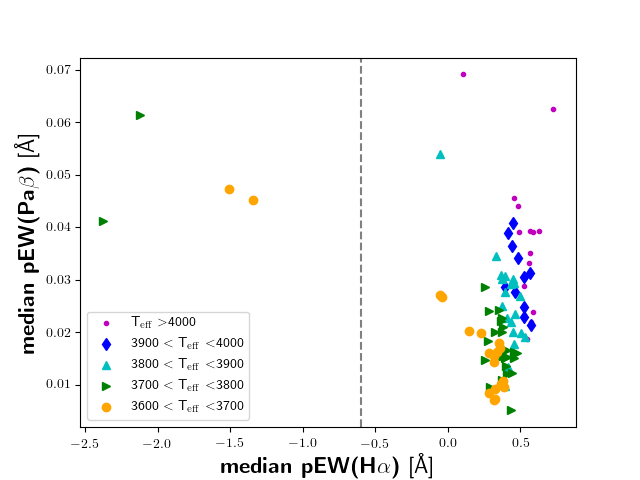}
\includegraphics[width=0.5\textwidth, clip]{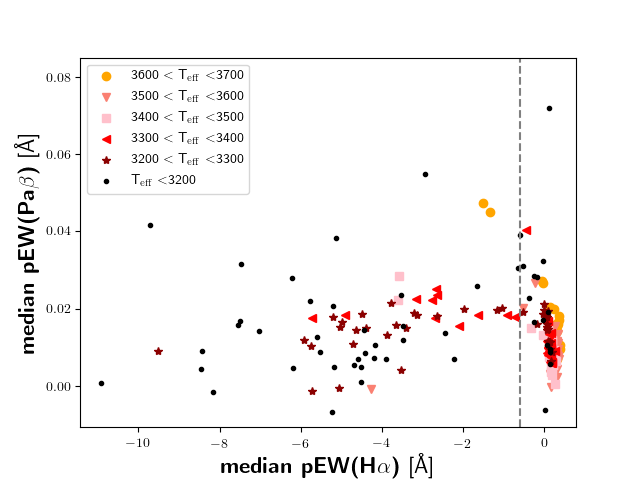}\\
	\caption{\label{fig:inactive} Median(pEW(Pa$\beta$)) shown in relation to 
	median(pEW(H$\alpha$)) with effective temperature of the stars colour-coded
	as shown in the legend. {\emph{Left:} Stars with high effective 
	temperature. \emph{Right:} Stars with low effective temperature. For
	better comparison we show stars with $3600 < T_{\rm eff} < 3700$\,K (orange circles)
	in both panels. The dashed vertical line marks the dividing line between active and inactive stars.} 
}
\end{center}
\end{figure*}

In Fig.~\ref{fig:inactive} we compare the median
observed pEWs of the H$\alpha$ and \pab\ lines of all 360 sample stars.  $T_{\rm eff}$
is shown  colour-coded in 100\,K intervals, to emphasise the temperature dependence of the \pab\ line. 
Looking at the stars with H$\alpha$ in absorption (pEW(H$\alpha$ > $-0.6$\,\AA; which we call inactive stars for the purpose of this study) in Fig.~\ref{fig:inactive},
an anti-correlation between $\mu$(pEW(Pa$\beta$)) and $\mu$(pEW(H$\alpha$)) can be noticed
for each effective temperature interval. To quantify this impression, we
calculated Pearson's correlation coefficients, $r$, for these samples, and obtained values
between $-0.42$ and $-0.91$ with p-values between 0.05 and $10^{-5}$, for 4000\,K $>$ $T_{\rm eff} >$ 3200\,K indicating
fair to highly significant correlations.
Only for the highest temperature interval with $r=-0.45$ and $p=0.13$ and for the lowest temperature
interval with $r=0.44$ and $p=0.08$ correlations are questionable (and would be positive for the lowest temperature stars). 
Thus, in general more absorption in H$\alpha$ is on average associated with less absorption in \pab\ in the stars with
H$\alpha$ in absorption for most temperatures. Since in our stellar sample typically
a larger pEW(H$\alpha$) is connected to less activity, the \pab\ line deepends for 
higher activity levels for the here considered inactive stars. This finding is in line
with the analysis by \citet{Patrick}.

Stars with H$\alpha$ in emission (which are called active stars traditionally) are only available in meaningful numbers in our sample
for $T_{\rm eff} < 3400$\,K and there is no comparable correlation for these; only
for the $ 3200 < T_{\rm eff} < 3300$\,K interval we find a fair correlation with $r=0.52$ and
$p=0.01$, the other two temperature intervals show no correlation with $r$ between $-0.03$ and 0.15 and
$p> 0.40$. Nevertheless, for stars with $T_{\rm eff} > $3600\,K, the pEW(\pab)
increases further for stars with H$\alpha$ in emission compared to the
pEW(\pab) of stars with H$\alpha$ in absorption. For stars with
$3200 < T_{\rm eff} < 3600$\,K and H$\alpha$ in emission, pEW(\pab) saturates
at the highest values found for stars with H$\alpha$ in absorption. For
stars with $T_{\rm eff} < 3200$\,K  saturation effects play a role 
for some of the stars, while others show higher or lower values of pEW(\pab).

Generally, for the coolest stars in our sample, the spread in pEW(\pab) is 
largest. Moreover, for these stars the most inactive stars with the lowest
pEW(\pab) values may be not present causing the mean apparent re-deepening of the
\pab\ line together with the additional absorption feature at 12821.4\,\AA. 
Nevertheless, as can be seen in Figs.~\ref{fig:teffpabeta} and~\ref{fig:inactive}, some of the coolest stars resume the trend to lower pEW(\pab).
These low values are found in more active stars and we interpret this as fillin-in of the line suggesting that the \pab\ line is very sensitive to the
pressure in the chromosphere.

\begin{figure}
\begin{center}
\includegraphics[width=0.5\textwidth, clip]{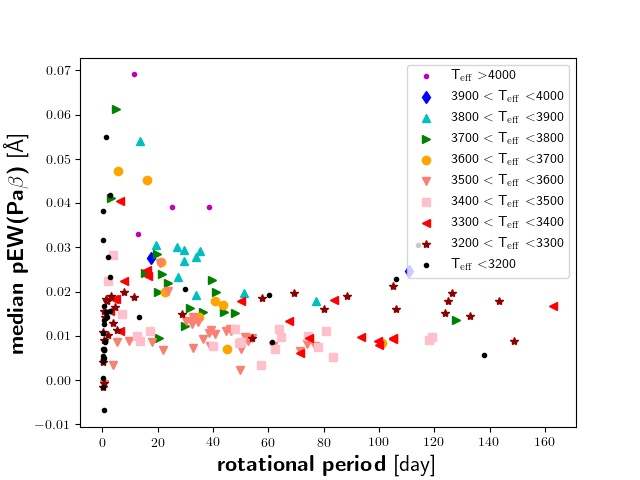}
        \caption{\label{fig:rotation} Median(pEW(Pa$\beta$)) shown in relation to
        the rotation period with effective temperature of the stars colour-coded
        as shown in the legend. 
}
\end{center}
\end{figure}

For 187 stars of our sample a rotational period is known. We therefore compare also
the median(pEW(Pa$\beta$)) to the rotational period for these stars (see Fig.~\ref{fig:rotation}). As discussed above, for the stars with higher effective temperature, which are generally more inactive, a deepening of the \pab\ line can be noticed towards shorter
rotation periods (i.\,e. higher activity levels). Only for the coolest stars 
deepening and fill-in or saturation is observed as one proceeds to shorter periods.

\subsubsection{Time series of individual stars}

\begin{figure}
\begin{center}
\includegraphics[width=0.5\textwidth, clip]{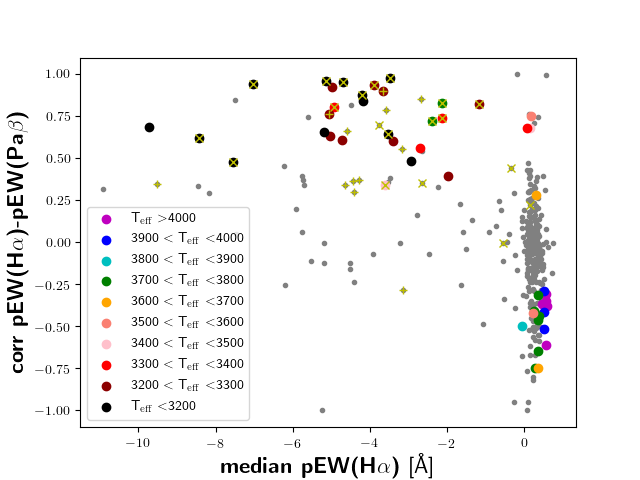}\\
\caption{\label{fig:correlation} Pearson correlation coefficient $r$ between
  pEW(H$\alpha$) and pEW(Pa$\beta$) shown as a function of $\mu$(pEW(H$\alpha$)).
  $T_{\rm eff}$ is colour coded as shown in the legend for stars with a significant
  Pearson correlation (p $<$ 0.005). Stars with automatically or visually
	found flares (see Sect.~\ref{sec:flares}) are marked with crosses
	and plusses, respectively.
}
\end{center}
\end{figure}

To study the relation between the pEWs measured in the H$\alpha$ and \pab\ lines in individual stars,
we also computed Pearson's correlation coefficients for the time series of pEW(H$\alpha$) and pEW(Pa$\beta$)
on a by-star basis and show the resulting values of $r$ in Fig.~\ref{fig:correlation}, where significant
results (p $<$ 0.005) are highlighted by colour.
Stars with H$\alpha$ in emission tend to show a significant positive
correlation between pEW(H$\alpha$) and pEW(Pa$\beta$),
while stars with H$\alpha$ in absorption more often show significant anti-correlation.
The latter cases are mostly  stars with $T_{\rm eff}>$3400\,K, while the former are usually cooler
stars, and one should keep in mind that our sample contains few stars with $T_{\rm eff}>$3400\,K
and H$\alpha$ in emission. Also, many of the stars showing positive 
correlations are affected by flaring activity in the Paschen lines
(see Sect.~\ref{sec:flares}), which often dominates the variability in pEW(Pa$\beta$)
and, thus, drives the correlation. 

Moreover, it is an interesting question, if detectable rotational modulation is imprinted on the
\pab\ time series. We therefore computed a generalized Lomb-Scargle periodogram \citep{Zechmeister2009}
for the \pab\ time series and accepted periods between 1.5 and 150\,days and a false-alarm probability smaller than 0.005 as significant rotational modulation. We found six stars
fulfilling these criteria, which we list in Table~\ref{tab:rot}. For three of these stars, there is no known rotation period. 
For one star a conflicting rotation period is known; for another star (J16343+571 / CM~Dra; spectral type M4.5\,V, eclipsing binary) 
we found a period of 2.4\,days, while a period of  about half this value of 1.27\,days is known for this
star for the mutual orbital period \citep{Doyle2000}. For the last star (J03133+047 / CD~Cet; spectral type M5.0\,V) we
found a period of 127.2 days, while a period of 126.2\,days was determined by \citet{Newton2016}. A more
detailed study found a period of $170^{+19}_{-38}$ days using photometry and about 134 days using spectroscopy
\citep{Bauer2020}. These findings show again
that the \pab\ line is sensitive to activity, but not as sensitive as other tracers. This may -- especially
for period search -- be partly caused by the problems of obtaining pEWs free of the influence of telluric
and airglow lines or the artefacts from their removal. Anyway,
rotational modulation in M dwarfs is traced best with photometric variations \citep{Irwin2011,West2015,SM16,DA19}.

\begin{table}
        \caption{\label{tab:rot} Detectable rotation in \pab. }
\footnotesize
\begin{tabular}[h!]{lrcl}
\hline
\hline
\noalign{\smallskip}

	star           & detected   & literature &ref   \\
   &  period     & period \\
   & [days] & [days]\\
\noalign{\smallskip}
\hline
\noalign{\smallskip}

	J00184+440 & 6.6 & ... \\
	J00403+612 & 119.3& ...\\
	J03133+047 & 127.2& 126.2 &New16 \\
	J04167$-$120 & 135.6 & ...\\
	J11511+352 & 94.0 & 22.8 $\pm$ 1.0 &DA19\\
	J16343+571 & 2.4 & 1.27 &Dev08\\
\noalign{\smallskip}
\hline

\end{tabular}
\normalsize
\tablebib{                                         
	DA19:~\citet{DA19}; Dev08:~\citet{Devor2008}; New16:~\citet{Newton2016}} 
\end{table}

\subsection{Flaring activity found in H$\alpha$ and the Paschen lines}
\subsubsection{Overview}\label{sec:flares}

\begin{figure}
\begin{center}
  \includegraphics[width=0.5\textwidth, clip]{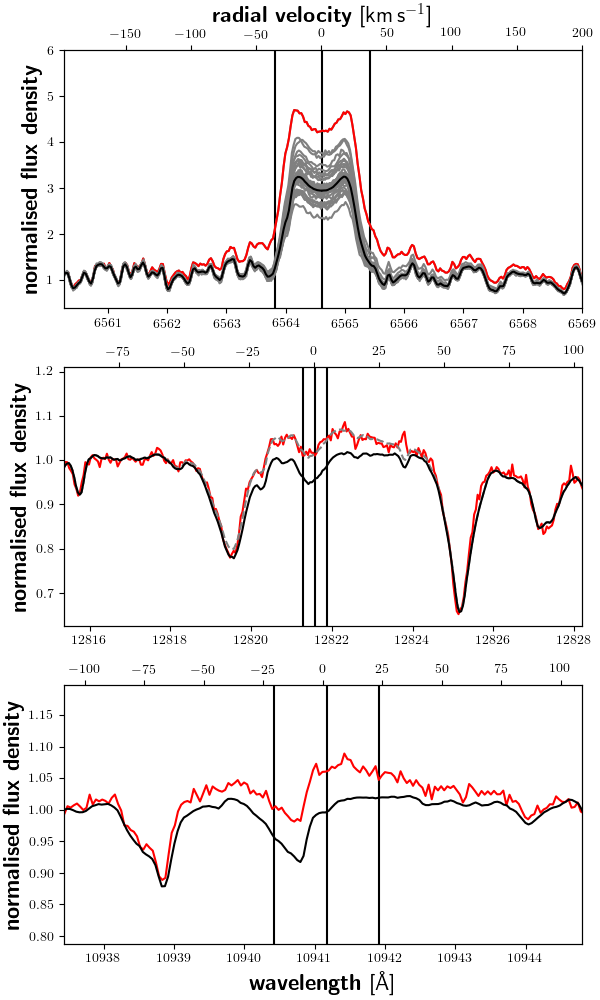}
  \caption{\label{fig:flareexample} Typical example of flaring activity: the
	M3.5\,V star J07319+362N / BL~Lyn.  {\emph Top:}
    The H$\alpha$ line. {\emph Middle:} The Pa$\beta$ line. {\emph Bottom:} The Pa$\gamma$
    line. Shown are all spectra of the star in grey (only for H$\alpha$ for clarity), and the median
    spectrum in black, while the flare spectrum is shown in red. For the \pab\ line we
    also show as grey dashed line a Gaussian fit of the flare excess flux density. The vertical lines
    denote the central wavelength of the respective line and the integration ranges of the
    pEWs. During flares showing \pab\ emission, the integration 
	range for the Paschen lines as well as the H$\alpha$ line is usually much too 
	small and can only be used for identification of these flares. In the shown example the full width at the line footpoints exceeds 5\,\AA\, for all 
	three shown lines. For more details, see Sects.~\ref{sec:pewmeasure} and \ref{sec:profiles}
}
\end{center}
\end{figure}

Applying the automatic flare search described in Sect.~\ref{sec:flaresearch}, we found 357 H$\alpha$ flares
in 153 stars, and 46 Pa$\beta$ flares in 30 stars.  
We summarize these and all the numbers in this Section in Table~\ref{tab:flsum}.
We examined all Pa$\beta$ flares by eye and removed 11 for which we found
airglow to remain a problem. The other stars all show the \pab\ line in emission. This excludes 
confusion with high amplitude rotational modulation, since \pab\ emission certainly involves high 
pressure, see also the discussion in Sect.~\ref{sec:Stark}. For these stars we 
automatically detect 15 stars with 24 Pa$\gamma$ flares and 15 stars with 24 Pa$\delta$
flares. These are not the same, though. While for most flares with Pa$\gamma$ emission,  Pa$\delta$
emission is also detected, there are six flares, where no Pa$\delta$ was detected. There are
another six flares, where Pa$\delta$ was detected despite no Pa$\gamma$ emission. For these
latter six flares the Pa$\delta$ detection is correct and Pa$\gamma$ was not detected due
to noise in some spectra of the three affected stars, which enlarges the MAD incorrectly and hinders the
automatic Pa$\gamma$  detection. Therefore we manually correct the number of Pa$\gamma$ flares to 30 flares in 18 stars.
As a typical example of the outcome of the automatic search, we show in 
Fig.~\ref{fig:flareexample},
the M3.5 star J07319+362N / BL~Lyn. Both, the Pa$\beta$ and the Pa$\gamma$ lines
can be seen in broad emission exceeding 5\,\AA. Other spectral features
in the region are still imprinted on the broad emission lines. The (about) 
Gaussian shape of these lines is revealed by spectral subtraction of the 
quiescent spectrum (see Sect.~\ref{sec:profiles}).

\begin{figure}
\begin{center}
\includegraphics[width=0.5\textwidth, clip]{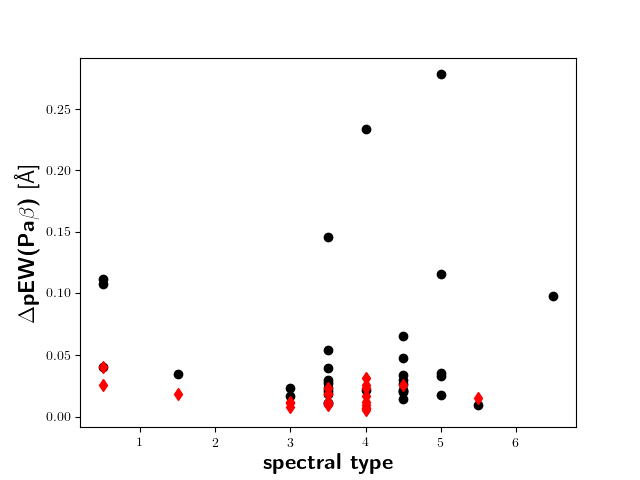}\\
        \caption{\label{scatterplot} 
          Deviation of the flare related pEW(\pab) from the median of the pEW(\pab) of the respective time series
          for the automatically detected flares (black dots) and the visually found flares (red diamonds).
	  We caution that $\Delta$pEW is systematically understimated, since the
	  integration width is not broad enough to cover the flaring line.
}
\end{center}
\end{figure}

An additional visual inspection of the stars with automatic \pab\ flare detections yielded another six small
Pa$\beta$ flares in four of the stars. 
Therefore, especially lower amplitude Pa$\beta$ flares may be missed
by the automatic detection, and we therefore screened our whole sample also by eye.
Thus, we identified 12 additional stars with 16
small Pa$\beta$ flares. We show an example spectrum of a visually found \pab\ flare
in Fig.~\ref{fig:notfound}. Generally, these visually found flares are comparable in strength to the smallest flares found by the automatic detection,
and the reasons why they were missed are manifold. One star, for instance,
shows a large red asymmetry, which shifts the Pa$\beta$ emission out of the integration range.
In other cases, the low number of spectra, large ranges of variability, and airglow or a combination thereof confound the search.
This altogether leaves us with 32 stars showing 57 Pa$\beta$ flares in comparison to
153 stars showing 357 H$\alpha$ flares. 
Since our flare classification relies on relative variation based on the MAD of the time series, we
  show in Fig.~\ref{scatterplot} also the absolute deviation of the flare related pEW(\pab) measurements compared
  to the median of pEW(\pab). We caution that these values are systematically underestimated, since our pEW
  integration range is not broad enoug to cover the whole line during the flare. Nevertheless, flares with
  $\Delta$pEW(\pab)$>0.03$\,\AA\, are typically detected automatically, while for smaller flares a
  non-detection by the search
algorithm gets more probable.
Furthermore, we note that some of the detected Paschen flares have been touched upon in the literature in the context of studies of
other lines \citep[]{asym, hevar}, but no detailed discussion of the specific properties of the Paschen lines
was provided there.

\begin{table}
        \caption{\label{tab:flsum} Summary of found flares. }
\footnotesize
\begin{tabular}[h!]{lcccc}
\hline
\hline
\noalign{\smallskip}

	method  &no. of         & no. of & no. of  & no. of \\  
	&\pab\    & Pa$\gamma$   & Pa$\delta$  & H$\alpha$ flares$^{a}$\\
               & flares$^{a}$ & flares$^{a}$ & flares$^{a}$ \\
\noalign{\smallskip}
\hline
\noalign{\smallskip}
	automatically &  46 (30) & 29 (20) & 27 (17) & 357 (153)\\
	after vis. exclusion  \\
	of false positives & 35 (20)& 24 (15) & 24 (15) \\
	after correction for \\
	noise in Pa$\gamma$&      & 30 (18) \\
	additionally visually\\ 
	found flares & 6 (4)$^{b}$ \\
	add. vis. found \\
	flares (whole sample) & 16 (12) & \\
	total & 57 (32) & 30 (18)& 24 (15) & 357 (153) \\

\noalign{\smallskip}
\hline

\end{tabular}
\normalsize
$^{a}$ The number of stars in which these flares are detected is given in parenthesis.\\
$^{b}$ In the stars with automatically \pab\ flare detection 
\end{table}

\subsubsection{Statistical properties of the Pa$\beta$ flares}

\begin{figure}
\begin{center}
\includegraphics[width=0.5\textwidth, clip]{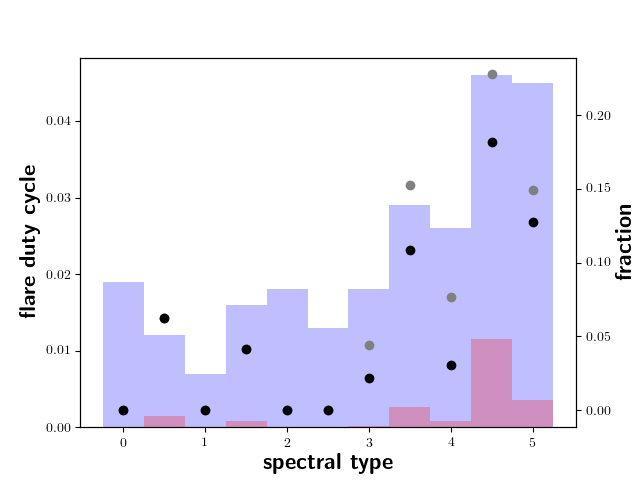}\\
        \caption{\label{fig:stats} 
    Flare duty cycle (flare time fraction, left y-axis) of H$\alpha$ (blue bars) and Pa$\beta$ flares (red bars) for
	automatically detected flares along with
    fractions of stars per spectral subtype (right y-axis) with automatically
	detected Pa$\beta$ flares (black dots) and all Pa$\beta$ flares (i.\,e. including visually
	found flares; grey dots). 
	The fractions of flaring stars for spectral types M0.5, M1.5, and M3.0 each correspond to a single star with a flare detection.
}
\end{center}
\end{figure}

The accumulated exposure time of all spectra considered here is 213.02~days.
Comparison with the total exposure time for spectra with automatically detected H$\alpha$ flares leads to
a ``flare duty cycle'' \citep{Hilton2010} of $2.26$\,\%. In contrast, the flare duty cycle for automatically
detected Pa$\beta$ flares, with all cases of airglow contamination excluded by visual inspection, is
only $0.19$\,\%, about an order of magnitude smaller. 

In Fig.~\ref{fig:stats},
we show the flare duty cycle  as a function of spectral subtype for
automatically detected H$\alpha$ and Pa$\beta$ flares, which increases toward later type stars for
both lines. For H$\alpha$ flares, such an increase was already described by \citet{Hilton2010}, who found,
however, lower duty cycles of 0.02\,\% for early M~stars and 3\,\% for late M~dwarfs in time resolved
spectra of the Sloan digital sky survey. 
We ascribe these different numbers to the different sensitivities and flare detection methods. 

Additionally, we show in Fig.~\ref{fig:stats}
the fraction of Pa$\beta$ flare stars as a function of the spectral subtype. The detected Pa$\beta$ flares
are clearly concentrated on stars of later spectral types, since only
three stars of type M3.0\,V and earlier show Pa$\beta$ flares, while the remaining 17 stars with automatically detected Pa$\beta$ flares
are of spectral type M3.5\,V and later.

Moreover, all but three stars with detected Pa$\beta$ flares clearly show H$\alpha$ in emission. The three stars with
H$\alpha$ absorption are J11476+786/GJ~445,
J02070+496 / G~173$-$037, and J23351$-$023 / GJ~1286, out of which the latter two are in a transition state, where H$\alpha$ is neither
in clear absorption nor emission. We show the flaring spectra of all of these three stars in Figs.~\ref{fig:abs1}, \ref{fig:abs2}, and \ref{fig:abs3}.

We also compare the flare duty cycle of all active stars, which we found to be
 4.7\,\% for H$\alpha$ and 1.0\,\% for \pab, while for the inactive stars it is 2.2\,\%
for H$\alpha$ and 0.03\,\% for \pab. These numbers are comparable to the values for
mid-type M dwarfs and early-type M dwarfs, since there are very few early-type M dwarfs
among the active stars, while the mid-type M dwarfs have many active stars among them.
Moreover, we compute the flare duty cycle only considering the stars flaring in H$\alpha$. Then the flare
duty cycle becomes 4.0\,\% for H$\alpha$ and 0.3\,\% for \pab.

\subsubsection{Stark broadening in H$\alpha$ for Pa$\beta$ flares}\label{sec:Stark}

Of the spectra exhibiting Pa$\beta$ emission, the
vast majority shows relatively symmetric H$\alpha$ line broadening.
In their analysis of line asymmetries,
\citet{asym} reported that red
asymmetries occur frequently, blue asymmetries are more rarely observed, and
symmetric line broadening is the most rarely observed variant, which is only about half as frequent as
red asymmetries. Therefore, we consider a chance finding unlikely and conclude that Pa$\beta$ emission is
likely coupled to the occurrence of symmetric broadening.
Like other authors such as \citet{Kowalski2017} and \citet{Wu2022}, we consider
Stark (pressure) broadening the most plausible explanation for the rather 
symmetric line profiles, which may alternatively be attributed to
turbulent broadening or an observational time integration effect, caused by the
blurring of a blue and a red asymmetry during the exposure.

\begin{figure}
\begin{center}
\includegraphics[width=0.5\textwidth, clip]{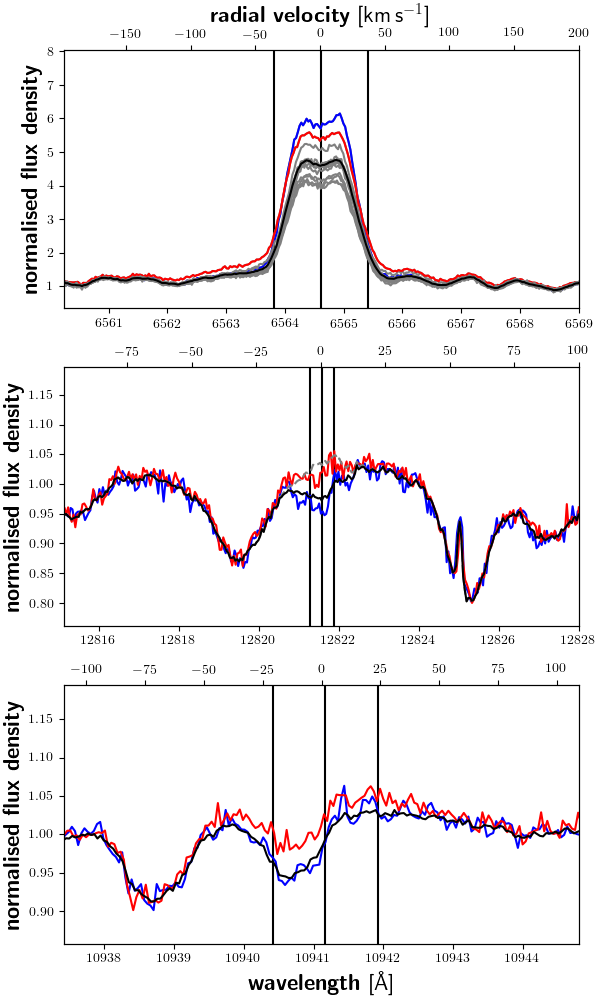}\\
        \caption{\label{nobroad} Same as  Fig. \ref{fig:flareexample} but for the M4.5\,V star J07558+833/GJ~1101. Two flares are marked as red and blue spectra, but only
	one leading to Pa line emission.}
\end{center}
\end{figure}

Stark broadening is a consequence of high pressure in the chromosphere and, therefore, is expected
to be associated with material showing larger collision rates, which lead to a larger population
of higher hydrogen excitation levels. Consequently, we attribute
the \pab\ line emission during flares to high pressures in the chromosphere
and lower transition region.
Notably, flares with 
comparable or even higher amplitudes in H$\alpha$ but no line broadening lead to
Paschen line emission as exemplified by the case shown in Fig.~\ref{nobroad}, where the
higher amplitude flare marked in blue does not show H$\alpha$ broadening and also no enhancement of
the flux in the Paschen lines. If anything, marginal excess absorption
may be present in Pa$\beta$
during this flare.
We caution, nevertheless, that Stark  broadening of the H$\alpha$ line
does not necessarily lead to Paschen line
emission. A case in point was presented by \citet{Paulson2006}, who reported on Stark broadening of
the Balmer lines during a flare on the M4.0 M~dwarf Barnard's star,
but did not detect Paschen line emission, although they covered Pa$\delta$ and higher.

\subsubsection{H$\alpha$ and Pa$\beta$ line profiles during flares}\label{sec:profiles}

Since the pEW(Pa$\beta$) is measured using a narrow integration band (see Table~\ref{tab:ew}), it captures only
a fraction of the flux if the line is broad. Likewise, the H$\alpha$ line also exceeds the integration width
used to obtain its pEW during many of the observed
flares. Therefore, the pEW values do not fully characterise the strengths of broad lines.

To obtain a better understanding of the line profiles during flares, we first
obtained excess spectra by subtracting the median spectrum from the flaring spectrum
and, subsequently, fitted
the resulting lines using Gaussians.
Specifically, we used a narrow and a broad Gaussian component for the
H$\alpha$ line as we did in \citet{asym} and a single Gaussian component for the \pab\ line.
We list the best-fit parameters of our model for each flare in Table~\ref{ew}
for the automatically detected flares and in Table~\ref{ew1} for all flares
found by visual inspection.

The two-Gaussian model reproduces the spectral line shape of H$\alpha$
fairly well. In particular, the two Gaussians
can account for mutual shifts of the broad and narrow component, which
is a measure of the line asymmetries.
For the Pa$\beta$ line, we used a single Gaussian model, which is
appropriate in most cases. There are eight Pa$\beta$ flares, where the fit of the excess flux in
Pa$\beta$ was not adequate. Six of 
these are visually found flares, where a combination of low amplitudes and
high width prevents a good fit. For some of the visually found flares also 
the broad component of H$\alpha$ cannot be fit for similar reasons. 
Additionally, there are a few examples where a single Gaussian  does not seem to be a suitable
model for the Pa$\beta$ line profile; these are marked in the Tables~\ref{ew} and \ref{ew1}.

With these fits we proceeded to investigate the correlation behaviour between the
H$\alpha$ and Pa$\beta$ line properties.
We found that the strength of the narrow H$\alpha$ component
is not correlated with that of the Pa$\beta$ line (Pearson's $r=0.23$, $p=0.11$).
Neither is the total strength of the narrow and broad H$\alpha$ components correlated with that of Pa$\beta$ 
($r=0.38$, $p=0.007$). However, the strength of the broad H$\alpha$ component is
correlated with that of the
Pa$\beta$ line (Pearson's $r=0.54$ and $p=7.7\cdot 10^{-5}$). Likewise, there are correlations
between the width
of the
broad H$\alpha$ component and that of the Pa$\beta$ line
($r=0.57$ and $p=2.2\cdot 10^{-5}$) and
the shift of the broad component of H$\alpha$  and the
line shift of Pa$\beta$ ($r=0.51$ and $p=0.0002$). This clearly shows that
the broad component of H$\alpha$ and the Pa$\beta$ emission are intimately related
during flares and that the emission originates most probably from the same material.

From the Gaussian fits also the luminosity of \pab\, $L_{\mathrm{\pab}}$ can be computed using
  PHOENIX photospheric models \citep{Husser2013} for $T_{\mathrm{eff}}$ and $\log g$ and  the radius of the
  respective star \citep{Schweitzer2019}. Using
  luminosities by \citet{Cifuentes2020} this can be converted into
  $\log L_{\mathrm{\pab}}/L_{\mathrm{bol}}$. We list these values in Tables~\ref{ew} and \ref{ew1}. 
  Values of $\log L_{\mathrm{\pab}}/L_{\mathrm{bol}}$ range from -4.03 to -8.16 but mostly concentrating between -5.5 and -6.5. We caution that these values strongly rely on theoretical assumptions and may therefore have large systematic errors.

\subsubsection{Asymmetries in Pa$\beta$ flares}

H$\alpha$ often exhibits asymmetric line profiles during flares, which can manifest
either as blue or red wing emission with various
velocity shifts and amplitudes (for asymmetries during flares on M dwarfs see \citet{asym}, and references therein; for the Sun see for example \citet{Berlicki2007}).
Asymmetric H$\alpha$ line spectra during flares are usually attributed to mass motions. Specifically, blue
asymmetries are thought to be caused by chromospheric evaporation during flare onsets
\citep{Li2022} or prominence eruption with possible coronal mass ejections \citep{Honda2018, Notsu2021}.
The origin of the red asymmetries is less certain and may, indeed, vary depending on whether the asymmetry is observed during the impulsive or the decay phase of the flare.  While red asymmetries may be caused by chromospheric condensations, mainly expected to happen
in the impulsive phase, in the decay phase they may be caused by coronal rain  \citep{Wu2022} or are associated with post flare loops \citep{Namizaki2023}.

Looking at the \pab\ line fits, we selected all spectra showing a shift of
more than 15\,km\,s$^{-1}\,(=0.65$\,\AA) and found two blue asymmetries and
one red asymmetry among the automatically detected flares and additionally
three red asymmetries among the visually found flares. Although these are small numbers, there seem to be more red than blue asymmetries, in agreement with
what \citet{asym} found for H$\alpha$ asymmetries. This seems to indicate again, that
the shifted \pab\ and H$\alpha$ emission originate in the same regions. The low number
of spectra with shifted \pab\ emission (compared to H$\alpha$) seem to indicate, that the pressure is usually not high enough
in these regions to produce a measureable amount of \pab\ emission.

\begin{figure*}
\begin{center}
  \includegraphics[width=0.5\textwidth, clip]{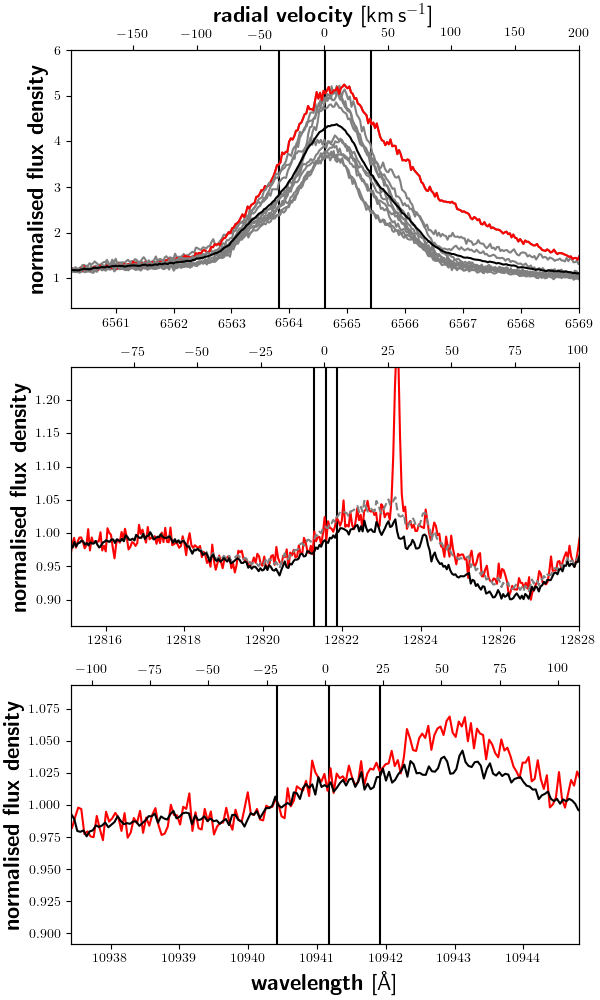}
  \includegraphics[width=0.5\textwidth, clip]{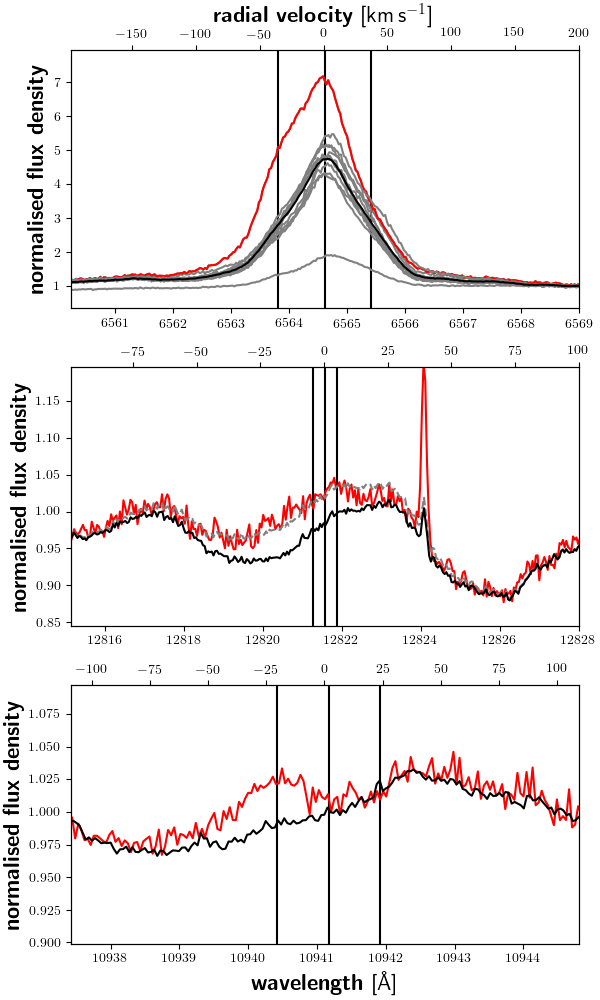}\\
  \caption{\label{fig:asym} Same as in Fig. \ref{fig:flareexample} but for the M4.0\,V star J01352$-$072 / Barta\,161\,12 (\emph{left}; not automatically found)
    and for the M4.0\,V star J22012+283 / V374~Peg (\emph{right}). Both stars are fast rotators with vsini = 59.8 and 36.9\,km\,s$^{-1}$, respectively.
  Both stars display large asymmetries in their H$\alpha$ lines and line shifts in the Pa$\beta$ line.}
\end{center}
\end{figure*}

\begin{figure}
\begin{center}
  \includegraphics[width=0.5\textwidth, clip]{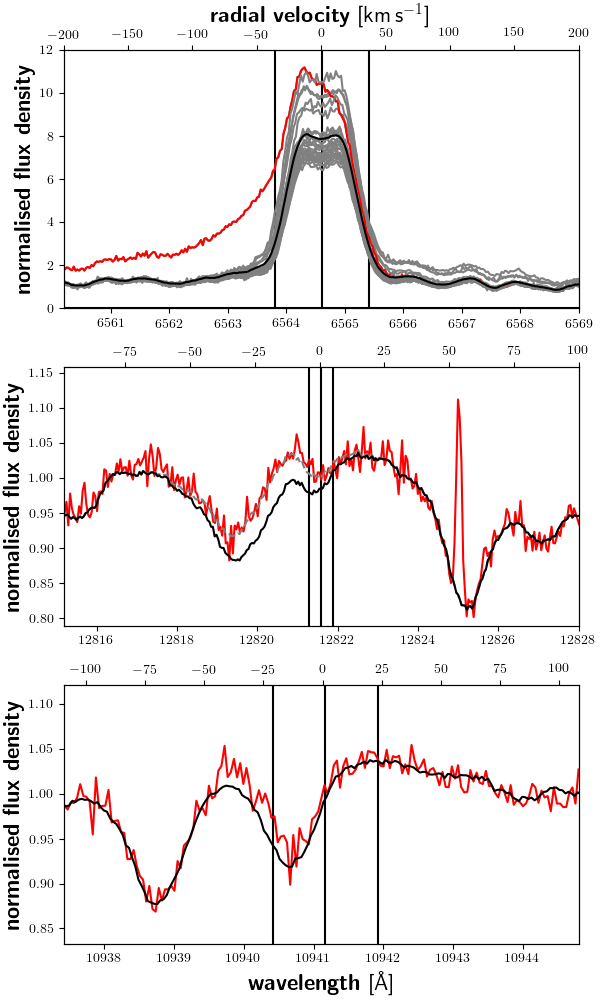}
  \caption{\label{fig:asym1} Same as in Fig. \ref{fig:asym} but for the M5.0\,V star J01033+623 / V388 Cas.
    The star is a moderately fast rotator with vsini = 10.5\,km\,s$^{-1}$. Again, the line shift in Pa$\beta$ corresponds to an asymmetry
  in the H$\alpha$ line. }
\end{center}
\end{figure}

Out of the six detected \pab\ asymmetries  we discuss here the two blue
asymmetries and the largest red asymmetry in more detail. 
While the latter belongs to  J01352$-$072 / Barta\,161\,12, the two blue asymmetries
occured on
J01033+623 / V388 Cas and J22012+283 / V374~Peg. 
For all three examples, the Pa$\beta$ lines show large shifts and the line profiles are consistent with
only the shifted material showing emission.
Therefore, we compared the line shifts (in velocity space) of the broad H$\alpha$ component to the line
shift of the Pa$\beta$ component.
For J01033+623 / V388\,Cas, both are shifted about $-30$\,km\,s$^{-1}$ compared to
the respective line centre. For J22012+283 / V374~Peg, we find a velocity of
$-8.7$\,km\,s$^{-1}$ for the broad component of H$\alpha$ and
$-16.4$\,km\,s$^{-1}$ for Pa$\beta$. For J01352$-$072 / Barta\,161\,12, we find $147.9$\,km\,s$^{-1}$ for H$\alpha$ and
$46.7$\,km\,s$^{-1}$ for
Pa$\beta$.
Given that the fit of very broad lines typically produces larger uncertainties on the central wavelength,
we consider the velocity shifts for the first two stars to be in agreement.
In the case of the red asymmetry, where the difference is about $100$\,km\,s$^{-1}$,
the broad H$\alpha$ profile may actually be composed of more than one component, which is not accounted for in
the modelling and could, thus, explain the difference.
We show all three examples in Figs.~\ref{fig:asym} and \ref{fig:asym1}.

There are more example of asymmetric H$\alpha$
line shapes in the spectra of the stars, which exhibit Pa$\beta$ flares at some point during the time series.
However, in these instances usually no Pa$\beta$ emission is detectable at all, neither at the
nominal wavelength nor at a shift. In these cases,
the densities in the moving material are likely too low to produce
Pa$\beta$ emission.

\subsubsection{Higher Paschen series lines}

\begin{figure}
\begin{center}
\includegraphics[width=0.5\textwidth, clip]{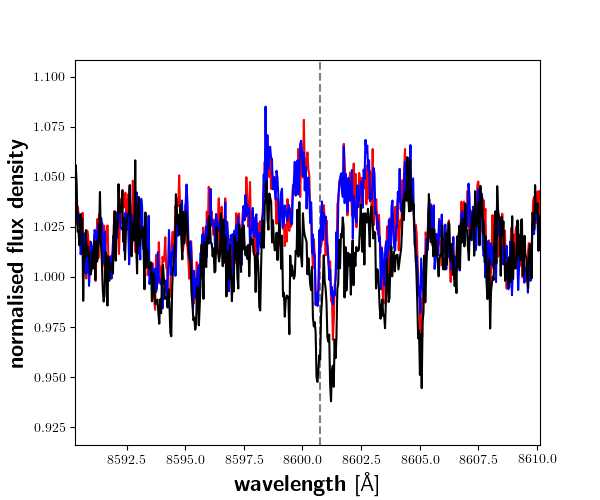}\\
        \caption{\label{fig:pa14a} Pa\,14 line for the M0.5\,V star J20451$-$313 / AU~Mic.
}
\end{center}
\end{figure}

We visually screened the spectra with detected flares also for
lines higher up in the Paschen series.
For seven stars, we could find Pa\,10 (at 9017.8\,\AA), Pa\,13 (at 8667.40\,\AA), and Pa\,14 (at 8600.75\,\AA) unambigiously.
Pa\,13, Pa\,15 (at 8547.73\,\AA), and Pa\,16 (at 8504.83\,\AA) 
are blended with the wings of the \ion{Ca}{ii} IRT at 8500.35, 8544.44, and 8664.52\,\AA.
Pa\,17 (at 8469.59\,\AA) is blended with two \ion{Ti}{i} absorption lines at 8469.474 and 8470.797\,\AA.
These Paschen lines are therefore hard to detect, especially next to the broad and highly variable 
\ion{Ca}{ii} IRT lines, which during flares usually show strong emission. 
We show the two flare spectra of J20451$-$313 / AU~Mic and the Pa\,14 lines in Fig. \ref{fig:pa14a} 
and the Pa\,14 line of J13536+776 / RX~J1353.6+7737 as second example
in Fig. \ref{fig:pa14}. We list all stars with detections of
Pa\,10 or higher in Table~\ref{tab:highpa}.

These higher Paschen lines have the potential to trace the gas conditions in
the chromosphere. A detailed chromospheric modelling with a stellar
atmosphere code would yield the best results. This was done for example using PHOENIX \citep{Hauschildt1999} by \citet{Hintz2020} for the \ion{He}{i} infrared
triplet. Such a modelling is beyond the scope of this paper and therefore we
stick here to a much easier and simpler analysis using the highest observed line, which was also used by \citet{Paulson2006} for a flare on Barnard's star
using hydrogen Balmer lines. The method of a pressure estimate using the highest resolved line is described in \citet{Kurochka1970}. For the flares, where we
detected only Pa\,10 as highest line, we argue that we are sensitivity limited.
Therefore, our highest resolved Paschen line is Pa\,13 or Pa\,14. Since for 
these high pressures broadening by the Doppler effect plays a minor role, the 
Stark effect dominates the broadening, which leads to a merging of the lines. Using
Equ.~9 from \citet{Kurochka1970} leads to $\log n_{e} \leqslant 14.0$. This value
compares well with the one found by \citet{Paulson2006} using the same method. It agrees also with the electron pressure found for the higher chromosphere by detailed flare modelling for a flare on CN~Leo \citep{Fuhrmeister2010}.


\begin{table}
        \caption{\label{tab:highpa} Flares with Pa lines higher than Pa 9 detected. }
\footnotesize
\begin{tabular}[h!]{lcccccccccc}
\hline
\hline
\noalign{\smallskip}

Karmn   & JD & highest line\\
\noalign{\smallskip}
\hline
\noalign{\smallskip}
J08298+267 &  2459177.625451389 & Pa 10 \\
J09161+018 & 2457712.658923611  & Pa 10\\
J11474+667 & 2457762.5463541667 & Pa 13\\
J13536+776 & 2458678.4093055557 & Pa 14 \\
J15218+209 & 2457950.498854167  & Pa 10\\
J20451$-$313 & 2458679.526851852  & Pa 14$^{*}$ \\
J20451$-$313 & 2458679.531608796  & Pa 14$^{*}$ \\
J22468+443 & 2457633.4671527776 & Pa 13\\

\noalign{\smallskip}
\hline

\end{tabular}\\
Notes: $^{*}$: These are consecutive spectra discussed in Sect.~\ref{sec:consecutive}.
\normalsize
\end{table}

\subsubsection{Consecutive Paschen line flares}\label{sec:consecutive}

There are three Paschen line flares with two consecutive spectra:
one flare on YZ Cet and two flares on AU Mic. In the case of YZ~Cet, the two
spectra were taken about one hour apart. For the flares on AU Mic, the temporal offsets are 
only 7 min and 14 min.
The observed evolution is diverse.
For the YZ Cet flare, the combined H$\alpha$ flux of the narrow and broad component
decayed by only ten percent within an hour, the Pa$\beta$ line flux approximately halved in the same time.
For the first AU Mic flare, the H$\alpha$ emission decayed by
about 15 percent in 7 minutes while the Pa$\beta$ emission stayed constant.
During the second flare, the H$\alpha$ emission decayed by about 80 percent in 14 minutes,
while the Pa$\beta$ emission dropped by about 95 percent.
Although the sample is small and the sampling sparse, we conclude tentatively
that the Pa$\beta$ emission during flares likely decays as fast as or faster
than the H$\alpha$ emission, but not significantly more slowly.

\subsubsection{Outstanding examples of Pa$\beta$ line flares}\label{sec:exc}
Among the identified  Pa$\beta$ flares, there are a number of exceptional examples.
We present here the stars with the largest Pa$\beta$ amplitude in our fit
(see Table~\ref{ew}). There are six flares with an amplitude larger than 1.0\,\AA\, belonging
to four stars. 
All four stars have rotational periods of less than 15 days and have generally high activity levels.

\begin{figure}
\begin{center}
\includegraphics[width=0.5\textwidth, clip]{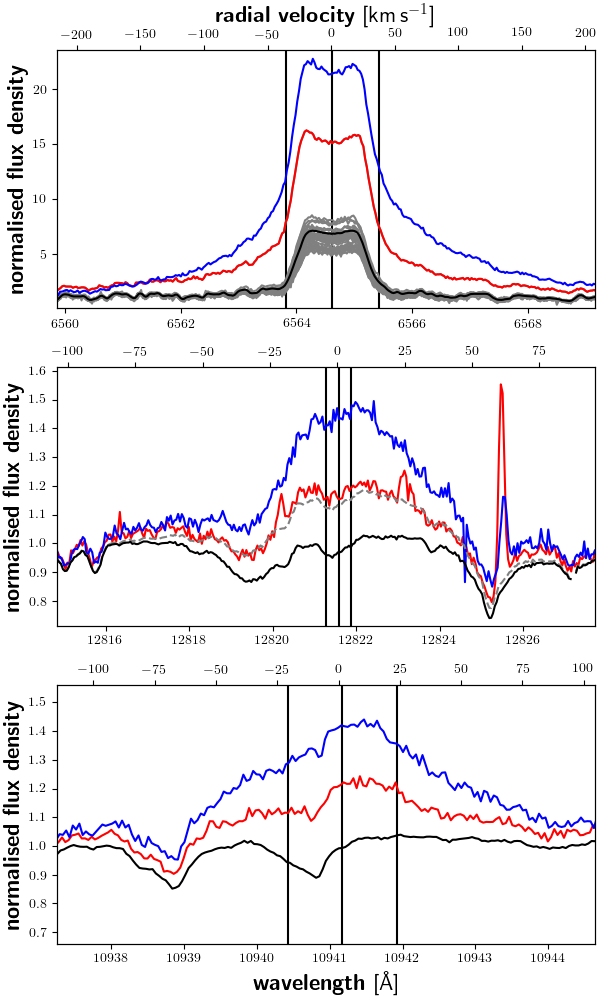}\\
        \caption{\label{fig:J11474}Same as in Fig. \ref{fig:flareexample} but for
        Pa$\beta$ flares on the M5.0\,V star J11474+667 / 1RXS~J114728.8+664405.
        Each coloured spectrum corresponds
        to one Pa$\beta$ flare.
}
\end{center}
\end{figure}

The star exhibiting the flare with the overall largest amplitude is the M5.0\,V star 
J11474+667 / 1RXS\,J114728.8+664405, which shows two Pa$\beta$ flares, with also the 
second flare having a considerable amplitude. We show the two flaring spectra
in Fig.~\ref{fig:J11474}. 

\begin{figure}
\begin{center}
\includegraphics[width=0.5\textwidth, clip]{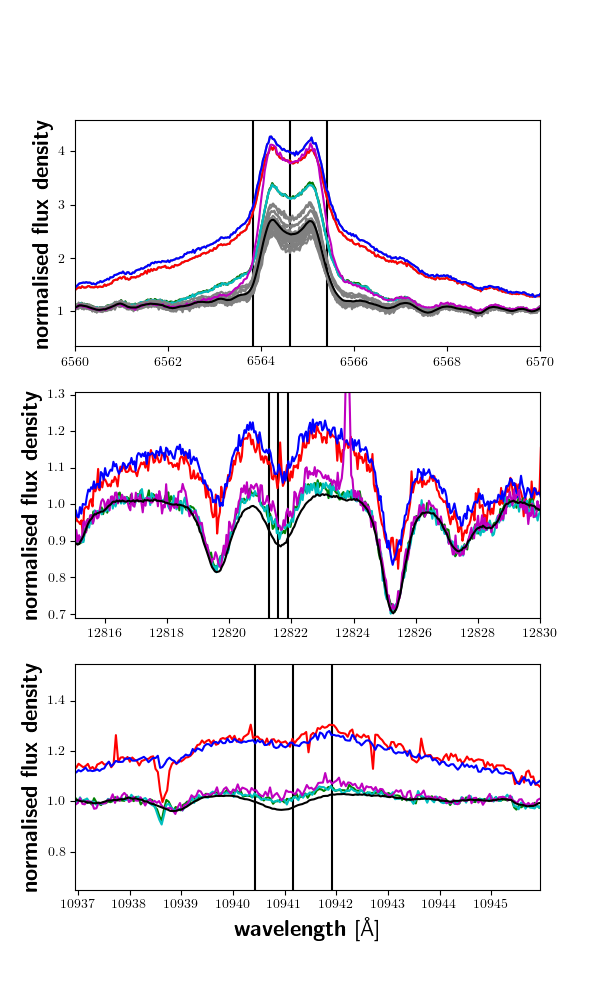}\\
        \caption{\label{fig:aumic}Same as in Fig. \ref{fig:flareexample} but for
        Pa$\beta$ flares on the M0.5\,V star J20451$-$313 / AU Mic. Each coloured spectrum corresponds
        to one Pa$\beta$ flare.
}
\end{center}
\end{figure}

The star with the second highest
amplitude is the young M0.5\,V star J20451$-$313 / AU~Mic, which has three out of four
automatically detected flares among the large amplitude flares. This is also the star
of
earliest subtype exhibiting a Pa$\beta$ flare and all three large Pa$\beta$ flares
of this star have the three broadest Pa$\beta$ lines found. We show the flaring
spectra  in Fig.~\ref{fig:aumic}.

Also J22468+443 / EV~Lac is an exceptional star. It has one flare among the large
Pa$\beta$ flares and another three automatically detected Pa$\beta$ flares. Considering
the three additional manually identified Pa$\beta$ flares, it is the star with the
largest number of Pa$\beta$ flares found (followed by J20451$-$313 / AU~Mic).
We show five (out of seven) flaring spectra of various strength in Fig.~\ref{fig:evlac}. 
The weakest of these flares shows Pa$\beta$ emission, but no detectable Pa$\gamma$ emission
while all other exhibit notable Pa$\gamma$ emission. 

The last star with a large amplitude Pa$\beta$ flare is the M4.0\,V star
J13536+776 / RX~J1353.6+7737. The spectra are shown in Fig. \ref{fig:J13536}.
J13536+776 / RX~J1353.6+7737 displays a second \pab\ flare, which is
quite a small one and cannot be fitted properly.

\section{Summary and conclusions}
In our study we analysed the Paschen lines, which are purely of chromospheric origin,
in a sample of 360 M dwarfs, which
provide together more than 19\,000 CARMENES spectra.  
We specifically used the pEW(H$\alpha$) and pEW(\pab) to characterise the behaviour
of the \pab\ line in non-flaring state along the M dwarf spectral sequence. 
We found, that on average the \pab\ line becomes more shallow for later spectral types 
until about spectral type M3.5; for even later spectral types the line re-deepens.
Comparing the pEWs of H$\alpha$ and \pab\ showed, that for inactive stars with
H$\alpha$ in absorption in a certain $T_{\rm eff}$ range, the median(pEW(H$\alpha$))
per star is anti-correlated to the median(pEW(\pab)). Only our hottest ($T_{\rm eff} > 4000$ K) and our coolest ($T_{\rm eff} < 3200$ K) temperature interval showed no
correlation. For the active stars with H$\alpha$ in emission, there is in contrast
only one $T_{\rm eff}$ interval, where a fair correlation between median(pEW(H$\alpha$))
and median(pEW(\pab)) could be found. 
Nevertheless, for  time series measurements of individual stars with H$\alpha$ in emission
we often found correlations between pEW(H$\alpha$) and pEW(\pab). On the other hand,
for time series
measurements of pEW(H$\alpha$) and pEW(\pab) we found an anti-correlation for many
stars with H$\alpha$ in absorption.
For both cases -- looking at the median values of the stars for comparing the stellar sample and also for looking at time series of individual stars -- we caution, that there are no stars with H$\alpha$ in emission for $T_{\rm eff}> 3400$K.

Regarding the flaring activity of the sample stars, we found 357 H$\alpha$ flares
in 153 stars in comparison to 30 (57) \pab\ flares in 18 (32) stars with the number
in brackets including flares found only by visual inspection. Out of the automatically
found \pab\ flares, 86\% and 69 \% also show Pa\,$\gamma$ and Pa\,$\delta$ in emission.
Even higher Pa lines could be found unambigously up to Pa\,14 for three flares (9\%). 
The detection of even higher Pa lines is hampered by their blending with the 
\ion{Ca}{ii} IRT or other stronger absorption lines. 
Since our pEW integration width is chosen mainly to identify flares for further characterization
  we applied Gaussian fitting to the \pab\ line. We demonstrate the quality of the Gaussian fit of the
  flare excess flux density by showing some of
these fits in Figs.~\ref{fig:flareexample}, \ref{nobroad}, \ref{fig:asym}, \ref{fig:asym1}, and~\ref{fig:J11474}.

Both, H$\alpha$ and
\pab\ flares are more often found in later spectral types (75\% of H$\alpha$ flares 
and 90\% of \pab\ flares are in stars with spectral type of M3.0\,V or later). The 'flare
duty cycle' (as a measure for the time fraction the star spends flaring) also increases
for later spectral types, as was found for H$\alpha$ already
by \citet{Hilton2010}. Moreover, the stars with \pab\ flares nearly all show H$\alpha$ in emission; only two show H$\alpha$ in a transition state from absorption to emission
and one star shows weak H$\alpha$ absorption outside the flares. Therefore, not
surprisingly, the stars with the most exceptional flares in amplitude, number and
width (which we show in Figs~\ref{fig:J11474}, \ref{fig:aumic}, \ref{fig:evlac} and \ref{fig:J13536}) are well known very active stars as J22468+443 / EV~Lac or J20451$-$313 / AU~Mic.
Additionally, we found some examples of asymmetries in the \pab\ lines during flares,
clearly more often associated with red asymmetries, than with blue ones.

Even more interestingly, \pab\ emission during flares seems to be coupled to high
densities, because almost all cases of \pab\ flaring occur, when H$\alpha$ exhibits
symmetric
broadening, which is indicative of Stark broadening \citep{Kowalski2017} and therefore 
high densities. Higher amplitude flares without Stark broadening typically do not lead
to \pab\ emission, but Stark broadening in H$\alpha$ also need not to necessarily lead to \pab\ 
emission.
As an indication of the strong coupling between the broad (Stark) component of the H$\alpha$
line and the \pab\ emission, we found a correlation between amplitude, width and shift
of these.
This sensitivity to chromospheric densities of the Pa lines deserves further
investigation. For this purpose,  dense time series of spectra covering a \pab\ flare are needed. We identified
here a number of promising candidates for such a project. These stars seem to show \pab\ flares more often than
the majority of M dwarfs.
Such flare observations would allow
to investigate, during which flare stage the \pab\ emission starts and if there is a time lag to
the reaction of H$\alpha$ like seen for other chromospheric lines in flare studies. Together with
dedicated chromospheric flare modelling this would lead to a better understanding of the
density variation during a flare.

\begin{acknowledgements}
  This publication was based on observations collected under the CARMENES Legacy+ project.
  CARMENES is an instrument at the Centro Astron\'omico Hispano en Andaluc\'ia (CAHA) at Calar Alto (Almer\'{\i}a, Spain), operated jointly by the Junta de Andaluc\'ia and the Instituto de Astrof\'isica de Andaluc\'ia (CSIC).
  The authors wish to express their sincere thanks to all members of the Calar Alto staff for their expert
  support of the instrument and telescope operation. The authors also thank the
    referee for the careful reading and the suggestions for improvement.
  CARMENES was funded by the Max-Planck-Gesellschaft (MPG), 
  the Consejo Superior de Investigaciones Cient\'{\i}ficas (CSIC),
  the Ministerio de Econom\'ia y Competitividad (MINECO) and the European Regional Development Fund (ERDF) through projects FICTS-2011-02, ICTS-2017-07-CAHA-4, and CAHA16-CE-3978, 
  and the members of the CARMENES Consortium 
  (Max-Planck-Institut f\"ur Astronomie,
  Instituto de Astrof\'{\i}sica de Andaluc\'{\i}a,
  Landessternwarte K\"onigstuhl,
  Institut de Ci\`encies de l'Espai,
  Institut f\"ur Astrophysik G\"ottingen,
  Universidad Complutense de Madrid,
  Th\"uringer Landessternwarte Tautenburg,
  Instituto de Astrof\'{\i}sica de Canarias,
  Hamburger Sternwarte,
  Centro de Astrobiolog\'{\i}a and
  Centro Astron\'omico Hispano-Alem\'an), 
  with additional contributions by the MINECO, 
  the Deutsche Forschungsgemeinschaft through the Major Research Instrumentation Programme and Research Unit FOR2544 ``Blue Planets around Red Stars'', 
  the Klaus Tschira Stiftung, 
  the states of Baden-W\"urttemberg and Niedersachsen, 
  and by the Junta de Andaluc\'{\i}a.
  We acknowledge financial support from the Agencia Estatal de Investigaci\'on (AEI/10.13039/501100011033) of the Ministerio de Ciencia e Innovaci\'on and the ERDF ``A way of making Europe'' through projects 
  PID2021-125627OB-C31,		
  PID2019-109522GB-C5[1:4],	
  and the Centre of Excellence ``Severo Ochoa'' and ``Mar\'ia de Maeztu'' awards to the Instituto de Astrof\'isica de Canarias (CEX2019-000920-S), Instituto de Astrof\'isica de Andaluc\'ia (SEV-2017-0709) and Institut de Ci\`encies de l'Espai (CEX2020-001058-M).
  This work was also funded by the Generalitat de Catalunya/CERCA programme and the Agencia Estatal de Investigaci\'on del Ministerio de Ciencia e Innovaci\'on (AEI-MCINN) under grant PID2019-109522GB-C53 and by the Deutsche Forschungsgemeinschaft under grant DFG SCHN 1382/2-1.
  This work made use of PyAstronomy \citep{Czesla2019pya}, which can be downloaded at {\tt https://github.com/sczesla/PyAstronomy}.
\end{acknowledgements}

\bibliographystyle{aa}
\bibliography{papers}

\appendix
\section{Comparison to PHOENIX models}\label{app:phoenix}

\begin{figure}
\begin{center}
\includegraphics[width=0.5\textwidth, clip]{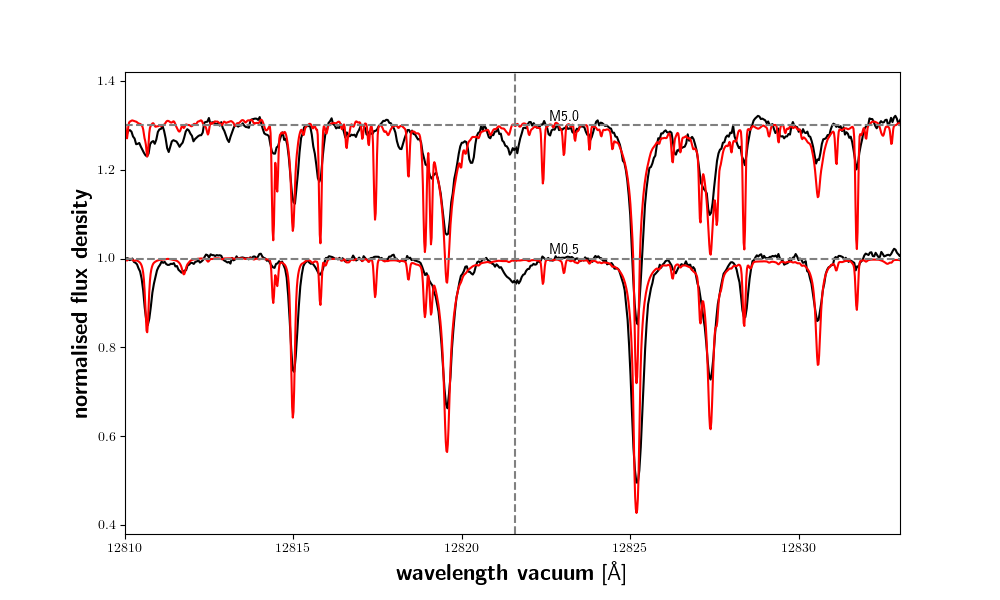}\\
\caption{\label{fig:phoenix} Spectral comparison to PHOENIX photospheric
  models for Pa$\beta$. Shown are the normalized spectra of the
  M0.5\,V star J02222+478 / BD+47~612 and the M5.0\,V star J18165+048 / G~140-051
  from Fig.~\ref{Mseq} as black solid line with an offset indicated by the grey dashed
  line. In red PHOENIX photospheric comparison spectra are shown. The vertical dashed line
  marks the position of the Pa$\beta$ line.
}
\end{center}
\end{figure}

Here we compare the spectral wavelength range around the Pa$\beta$ line of the
M0.5\,V star J02222+478 / BD+47~612 and the M5.0\,V star J18165+048 / G~140-051 (both also shown in Fig.~\ref{Mseq}) 
to PHOENIX purely photospheric
models from the library of \citet{Husser2013}. PHOENIX is a stellar atmosphere code \citep{Hauschildt1999},
which is widely used to compute photospheric stellar models and their synthesised
stellar spectra and has been applied to CARMENES spectra to establish stellar parameters
 \citep{Passegger2018, Schweitzer2019, Cifuentes2020, Marfil2021}.
We use a model with $T_{\rm eff}$=3900\,K
for the M0.5 star, a model with $T_{\rm eff}$=3200\,K for the M5.0 star and $\log g$=5.0
in both cases. \citet{Marfil2021} listed $T_{\rm eff}=3894\pm$11\,K and 3240$\pm$36\,K and
$\log g=4.99\pm$0.09 and 4.97$\pm$0.13, for the two stars, respectively. 
In Fig.~\ref{fig:phoenix} the generally good resemblance for both spectra can be seen.
Although there are some weaker lines in the PHOENIX spectra which are not seen in the
observed spectra and vice versa, the stronger atomic lines match quite well. However,
the Pa$\beta$ line is not present in the PHOENIX spectra, making it evident, that it is
a purely chromospheric line.

\section{Further examples of Pa$\beta$ line flares}\label{app:flare}

\begin{figure}
\begin{center}
  \includegraphics[width=0.5\textwidth, clip]{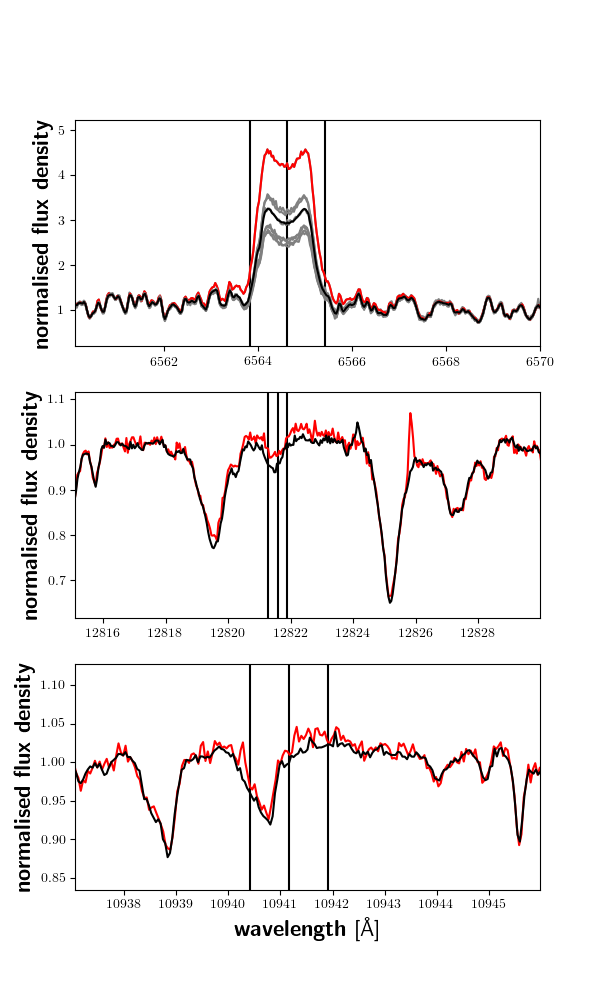}\\
  \caption{\label{fig:notfound} Same as in Fig. \ref{fig:flareexample} but for a Pa$\beta$ flare not found automatically for the M4.0\,V star J12428+418 / G~123-055.
    There are only few spectra available for the star and the variation is high, which prevents the program to
    automatically detect the relatively small flare.}
\end{center}
\end{figure}

\begin{figure}
\begin{center}
  \includegraphics[width=0.5\textwidth, clip]{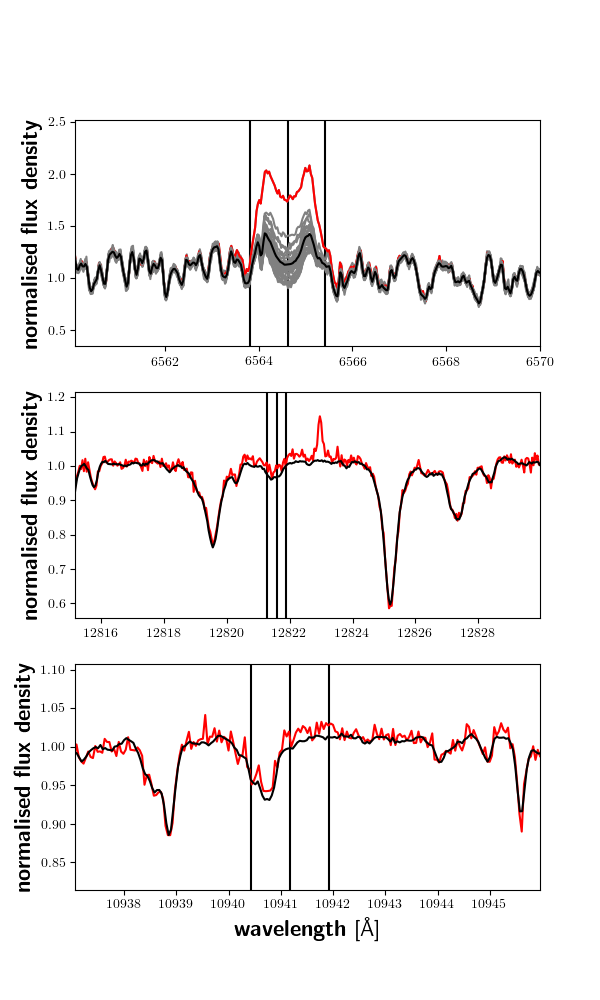}\\
  \caption{\label{fig:abs1} Same as in Fig. \ref{fig:flareexample} but for a Pa$\beta$ flare for the M3.5\,V star J02070+496 / G~173-037,
    which has H$\alpha$ in a transition state between absorption and emission.
    }
\end{center}
\end{figure}

\begin{figure}
\begin{center}
  \includegraphics[width=0.5\textwidth, clip]{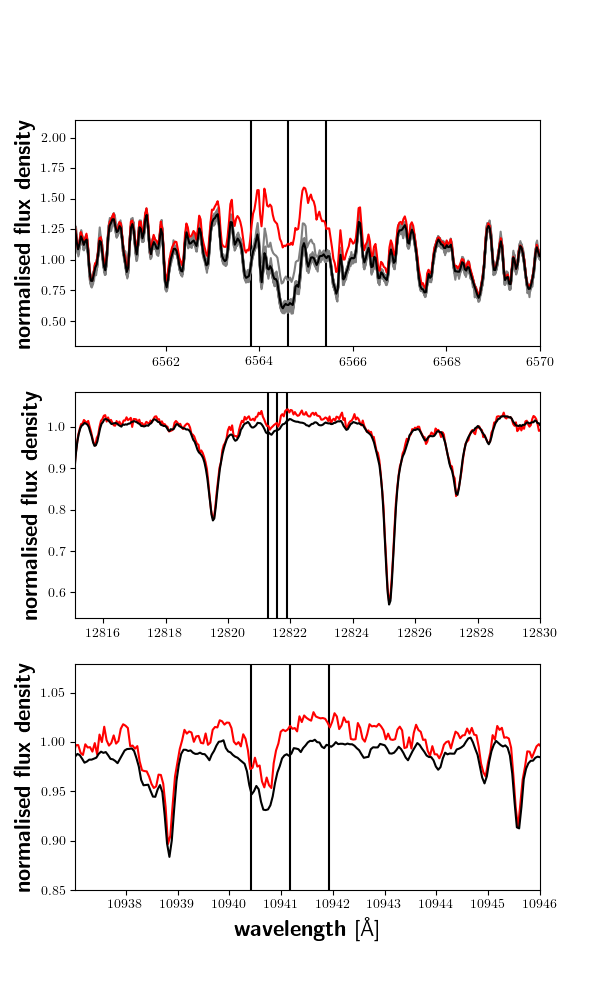}\\
  \caption{\label{fig:abs2} Same as in Fig. \ref{fig:flareexample} but for a Pa$\beta$ flare for the M3.5\,V star J11476+786 / GJ~445,
    which has H$\alpha$ in weak absorption.
    }
\end{center}
\end{figure}

\begin{figure}
\begin{center}
  \includegraphics[width=0.5\textwidth, clip]{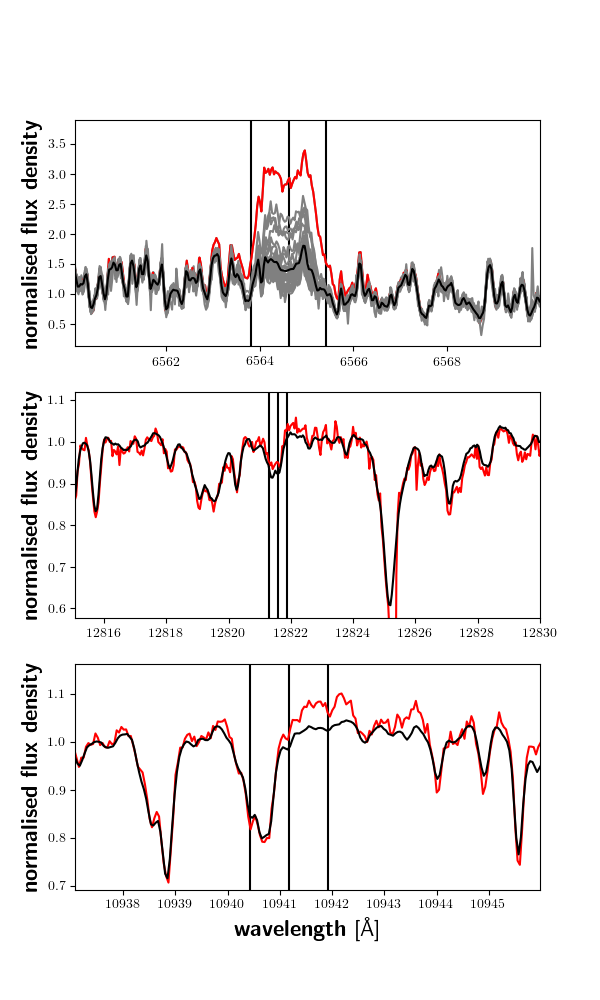}\\
  \caption{\label{fig:abs3} Same as in Fig. \ref{fig:flareexample} but for a Pa$\beta$ flare for the M5.5\,V star J23351$-$023 / GJ~1286,
    which has H$\alpha$ in a transition state between absorption and emission.
    }
\end{center}
\end{figure}

\begin{figure}
\begin{center}
\includegraphics[width=0.5\textwidth, clip]{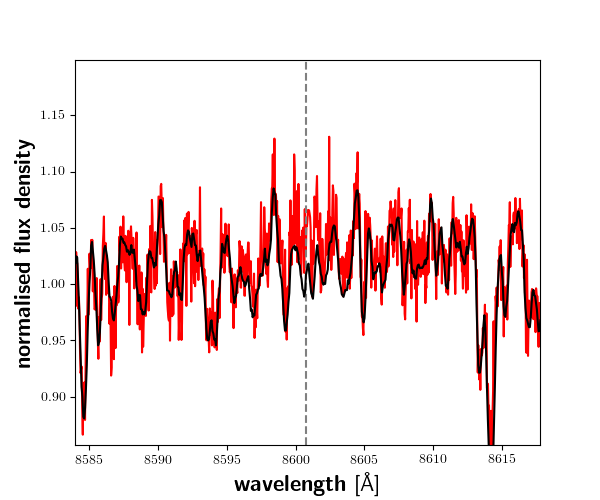}\\
        \caption{\label{fig:pa14} Pa\,14 line for the M4.0\,V star J13536+776 / RX~J1353.6+7737.
}
\end{center}
\end{figure}

\begin{figure}
\begin{center}
\includegraphics[width=0.5\textwidth, clip]{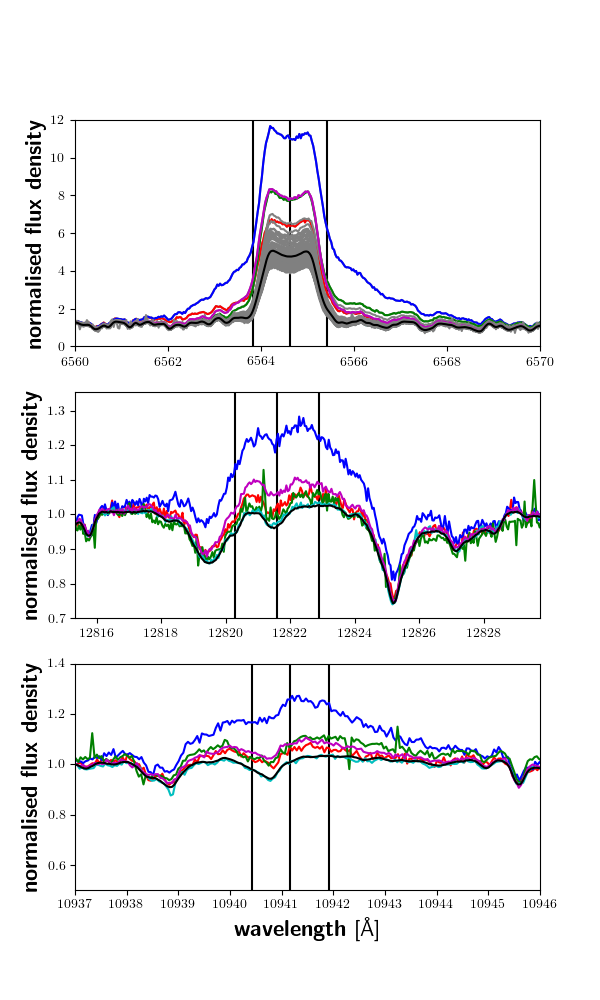}\\
        \caption{\label{fig:evlac}Same as in Fig. \ref{fig:flareexample} but for
        Pa$\beta$ flares on the M3.5\,V J22468+443 / EV~Lac. Each coloured spectrum corresponds
        to one Pa$\beta$ flare. The weakest flare is marked in cyan and has no Pa$\gamma$ emission any more.
}
\end{center}
\end{figure}

\begin{figure}
\begin{center}
\includegraphics[width=0.5\textwidth, clip]{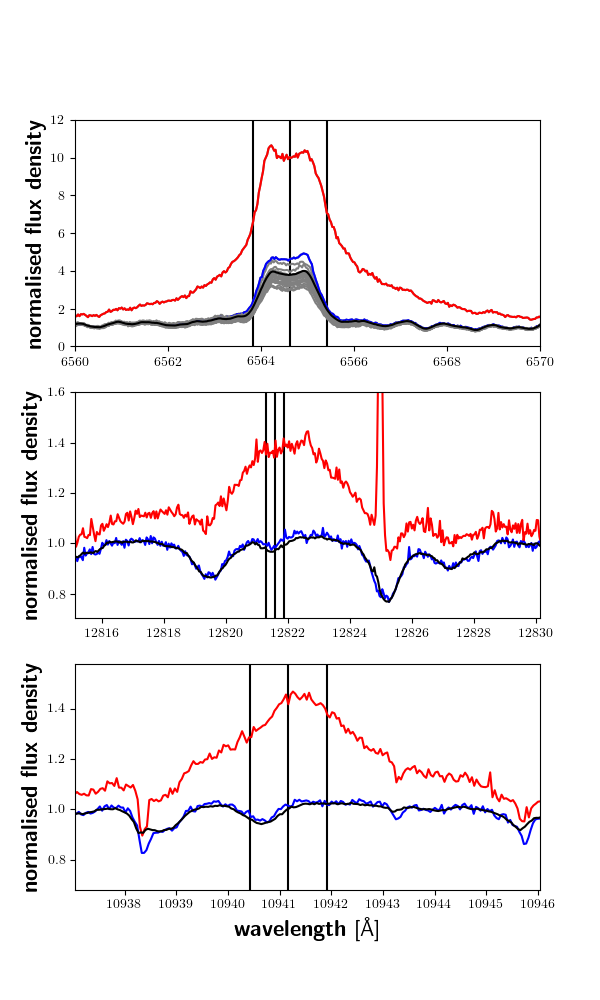}\\
        \caption{\label{fig:J13536}Same as in Fig. \ref{fig:flareexample} but for
	  Pa$\beta$ flares on the M4.0\,V star J13536+776 / RX~J1353.6+7737.
          Each coloured spectrum corresponds
        to one Pa$\beta$ flare.
}
\end{center}
\end{figure}

We show here further examples of Pa$\beta$ line flares. 
In Fig. \ref{fig:notfound} we show an example of a flare not found
by our automatic detection, but only by visual inspection.
In Figs. \ref{fig:abs1}, \ref{fig:abs2}, and \ref{fig:abs3} we show
the flare spectra for the three stars with H$\alpha$ not in clear emission.
We show a further example of the Pa\,14 line  in Fig.~\ref{fig:pa14}.
In Figs.~ \ref{fig:evlac},
and \ref{fig:J13536} we show the outstanding flares, which we described in Sect.~\ref{sec:exc}.

\clearpage
\section{Gaussian fitting of the H$\alpha$ and Pa$\beta$ lines affected by flaring}\label{app:ewfit}

The fitting parameters of the Gaussian line fitting of the H$\alpha$ and Pa$\beta$ lines, which are affected by flaring, can be found in the Tables \ref{ew} and \ref{ew1} for the
automatically and manually found flares, respetively. H$\alpha$ is fitted with
a broad and a narrow Gaussian component, while we fit the Pa$\beta$ line with only one Gaussian component as is appropriate for most of the lines. Nevertheless, for
some lines no good fit could be established. 

\begin{sidewaystable*}
        \caption{\label{ew} Gaussian fitting parameters for the H$\alpha$ and Pa$\beta$
        line for automatically detected Pa$\beta$ flares. }
\footnotesize
\begin{tabular}[h!]{llllcrrcrrcrrrrr}
\hline
\hline
\noalign{\smallskip}

	Karmn           & Name   & Spec.&JD  & $\lambda_ {\rm c}$  & Area & $\sigma$ &  $\lambda_ {\rm c}$   & Area & $\sigma$ &  $\lambda_ {\rm c}$   & Area & $\sigma$ & log($L_{\mathrm{\pab}}/L_{\mathrm{bol}}$) & $\Delta$pEW(\pab) &shown in \\
                &     &type& -2450000&    &       &          & &   &   & \\
		&     & &[day] & [\AA] & [\AA] & [\AA] & [\AA] & [\AA] & [\AA] & [\AA] & [\AA] & [\AA] & & [\AA] & Fig.\\
\noalign{\smallskip}
\hline
\noalign{\smallskip}
	J01019+541 & G 218-020 &M5.0\,V&7761.311  & 6564.70 & 7.40 & 0.59 & 6565.16 & 3.1 & 2.6 & 12822.1 & 0.18 & 1.90 & -6.64 &0.035\\
	J01033+623 & V388 Cas & M5.0\,V&7814.315 & 6563.87 & 5.66 & 0.75& 6561.48 & 5.02 & 1.98 & 12820.20 & 0.12 & 1.05 & -5.76 & 0.017 &\ref{fig:asym1}\\
	J01125$-$169 & YZ Cet & M4.5\,V&7959.673 &6564.60 & 1.42 & 0.60 & 6565.70 & 0.55 & 1.57 & ... & ... & ... & ...& 0.014\\ 
	J01125$-$169 & YZ Cet & M4.5\,V&8451.355& 6564.61 & 2.99 & 0.55 & 6564.71 & 0.54 & 2.00 & 12821.50 & 0.12 & 1.12 & -6.02 &0.022\\
	J01125$-$169 & YZ Cet & M4.5\,V&8474.265 & 6564.61 & 1.48 & 0.64 & 6564.55 & 2.13 & 2.81 & 12821.50 & 0.12 & 1.38 & -6.02& 0.021\\ 
	J01125$-$169 & YZ Cet & M4.5\,V&8487.277 & 6564.63 & 3.65 & 0.64 & 6566.72 & 0.46 & 0.84 & 12821.50 & 0.10 & 1.07 & -6.10 &0.020\\
	J01125$-$169 & YZ Cet & M4.5\,V&8493.305 &  6564.59 & 7.47 & 0.47 & 6564.62 & 2.36 & 1.71 & 12821.60 & 0.20 & 0.97 & -5.80 & 0.047\\
	J01125$-$169 & YZ Cet & M4.5\,V&8493.348  & 6564.60 & 6.14 & 0.46 & 6565.24 & 2.61 & 1.56 & 12821.60 & 0.09 & 0.67 & -6.14& 0.030\\
	J02070+496 & G 173-037&M3.5\,V&8499.444 & 6564.62 & 0.90 &  0.53 & 6566.40 & 0.06 & 1.4  & 12822.30 & 0.17 & 1.80 & -6.44 & 0.011& \ref{fig:abs1}\\
	J07319+362N&  BL Lyn& M3.5\,V&7449.384  & 6564.59 & 2.23 & 0.73 & 6565.61 & 1.49 & 2.10 & 12821.60 & 0.21 & 1.40 & -6.94 & 0.039 &\ref{fig:flareexample}\\ 
	J07446+035 & YZ CMi & M4.5\,V&7788.475  & 6564.56 & 2.72 & 0.89 & 6564.21 & 5.00 & 4.14 & 12821.60 & 0.75 & 3.20 & -4.29& 0.034\\
	J08298+267 & DX Cnc &M6.5\,V&9177.625  & 6564.70 & 30.0 & 0.52 & 6565.07 & 10.37 &1.57 & 12821.60$^{+}$ & 0.39 & 0.97 & -4.03& 0.098\\ 
	J10196+198 & AD Leo & M3.0\,V&8209.469 & 6564.62 & 1.92 & 0.60 & 6566.83 & 0.15 & 0.92 & 12821.30 & 0.07 & 1.08 & -8.16 & 0.017\\
	J10196+198 & AD Leo & M3.0\,V&9727.355 & 6564.63 & 2.09 & 0.60 & 6565.39 & 0.37 & 1.88 & 12821.90 & 0.19 & 2.04 & -7.73& 0.023\\
	J11474+667 & 1RXS J11472+66$^{a}$& M5.0\,V&7762.546 & 6564.60 & 13.58& 0.72 & 6564.33 & 15.83& 3.70 & 12821.60$^{+}$ & 0.95 & 2.05 & -5.99 & 0.116&\ref{fig:J11474}\\ 
	J11474+667 & 1RXS J11472+66$^{a}$& M5.0\,V&8852.717 &  6564.61 & 22.02& 0.75 & 6565.28 & 25.0 & 2.36 & 12821.60$^{+}$ & 2.20& 1.90 & -5.62 & 0.278&\ref{fig:J11474}\\ 
	J11476+786 & GJ 445 &M3.5\,V &8845.663 & 6564.61 & 0.72 & 0.64 & 6564.71 & 0.58 & 2.58 & 12821.80 & 0.10 & 2.08 & -6.14 & 0.011&\ref{fig:abs2}\\ 
	J13536+776 & RX J1353.6+7737&M4.0\,V&8678.409 & 6564.69 & 10.28& 0.86 & 6565.03 & 14.56& 3.11 & 12822.20$^{+}$ & 2.14 & 2.19 & -4.25 & 0.234&\ref{fig:J13536}\\ 
	J13536+776 & RX J1353.6+7737&M4.0\,V&8877.716* & 6564.57 & 1.17 & 0.52 &  ...     & ... &  ...    & ... & ... & ... & ...& 0.012& \ref{fig:J13536}\\
	J15218+209 & OT Ser & M1.5\,V &7752.704* & 6564.64 & 0.46 & 1.14 & 6564.54 & 1.27 & 4.08 & 12821.60 & 0.15 & 1.79 & -5.09& 0.018\\
	J15218+209 & OT Ser & M1.5\,V &7950.498  & 6564.62 & 1.37 & 0.76 & 6564.86 & 1.12 & 2.37 & 12821.60 & 0.39 & 2.18 & -4.69& 0.034\\
	J18075$-$159 & GJ 1224&M4.5\,V &8700.383 & 6564.62 & 7.11 & 0.61 & 6565.35 & 1.55 & 1.96 & 12821.70 & 0.38 & 1.55 & -6.10&0.065\\
	J18482+076 & G 141-036 & M5.0\,V&7631.455 & 6564.55 & 5.02 & 0.70 & 6564.93 & 3.15 & 1.94 & 12821.60 & 0.18 & 1.30 & -5.81& 0.033\\
	J18498$-$238 & V1216 Sgr &M3.5\,V&8033.285  & 6564.56 & 0.70 & 0.48 & ...       & ... & ... & ... & ...& ...          & ...    & 0.023\\
	J18498$-$238 & V1216 Sgr &M3.5\,V&8264.646 & 6564.63 & 1.02 & 0.57 & 6564.66 & 0.38 & 1.91 & 12821.60 & 0.09 & 1.58 & -6.52&0.020\\
	J18498$-$238 & V1216 Sgr &M3.5\,V&8300.556 & 6564.65 & 4.98 & 0.56 & 6564.92 & 1.70 & 1.76 & 12821.60 & 0.20 & 1.60 & -6.18&0.029\\
	J20451$-$313 & AU Mic & M0.5\,V&8679.526  & 6564.67 & 4.87 & 1.93 & 6564.53 & 5.88 & 5.78 & 12821.60 & 1.73 & 4.06 & -5.13 &0.107&\ref{fig:aumic}\\
	J20451$-$313 & AU Mic & M0.5\,V&8679.531  & 6564.69 & 2.94 & 1.60 & 6564.56 & 6.31 & 3.96 & 12821.60 & 1.77 & 4.12 & -5.13 &0.112&\ref{fig:aumic}\\
	J20451$-$313 & AU Mic & M0.5\,V&8680.522 & 6564.72 & 2.53 & 1.43 & 6564.41 & 7.79 & 3.70 & 12821.50 & 2.18 & 4.48 & -5.04 &0.040&\ref{fig:aumic}\\
	J20451$-$313 & AU Mic & M0.5\,V&8680.532  & 6564.91 & 0.47 & 0.60 & 6564.57 & 1.53 & 1.49 & 12821.40 & 0.12 & 1.42 & -6.30 &0.026&\ref{fig:aumic}\\
	J20451$-$313 & AU Mic & M0.5\,V& 9161.271* & 6564.87 & 0.46 & 0.59 & 6564.61 & 1.47 & 1.48 & 12821.60 & 0.10 & 1.18 & -6.37& 0.040\\ 
	J22012+283 & V374 Peg&M4.0\,V&7754.323  & 6564.24 & 3.72 & 0.62 & 6564.43 & 1.53 & 2.65 & 12820.90 & 0.22 & 2.32 & -6.06 & 0.022&\ref{fig:asym}\\
	J22231$-$176 & LP 820-012& M4.5\,V&7586.616 & 6564.63 & 2.30 & 0.65 & 6564.98 & 3.93 & 1.72 & 12821.90 & 0.12 & 1.34 & -6.77& 0.026\\
	J22468+443 & EV Lac & M3.5\,V&7626.537*  &  6564.59 & 1.74 & 0.55 & ... & ... & ... & ...& ...& ...  & ...&0.011\\
	J22468+443 & EV Lac & M3.5\,V&7632.628 & 6564.57 & 2.42 & 0.72 & 6563.81 & 2.35 & 1.97 & 12821.60 & 0.22 & 1.93 & -6.01 & 0.027&\ref{fig:evlac}\\
	J22468+443 & EV Lac & M3.5\,V&7633.467 & 6564.61 & 7.67 & 0.68 & 6565.07 & 9.92 & 1.75 & 12821.60 & 1.16 & 1.83 & -5.29 & 0.146&\ref{fig:evlac}\\
	J22468+443 & EV Lac & M3.5\,V&7647.373* & 6564.72 & 0.51 & 0.72 & 6565.36 & 1.29 & 3.35 & 12821.50 & 0.10 & 1.07 & -6.35& 0.018\\
	J22468+443 & EV Lac & M3.5\,V&7650.536 & 6564.59 & 4.38 & 0.59 & 6566.30 & 2.81 & 1.67 & 12821.60 & 0.10 & 1.03 & -6.35 & 0.017&\ref{fig:evlac}\\
	J22468+443 & EV Lac & M3.5\,V&7931.662* & 6564.63 & 1.66 & 0.68 & 6565.26 & 0.44 & 1.84 & 12821.60 & 0.18 & 1.97 & -6.10& 0.023& \ref{fig:evlac}\\
	J22468+443 & EV Lac & M3.5\,V&8032.429 & 6564.55 & 4.98 & 0.62 & 6565.20 & 1.46 & 1.96 & 12821.50 & 0.33 & 1.27 & -5.83 & 0.054&\ref{fig:evlac}\\
	J23351$-$023 & GJ 1286& M5.5\,V&8080.358& 6564.61 & 2.27 & 0.57 & 6563.87 & 0.41 & 2.28 & 12821.70 & 0.03 & 0.50 & -6.65& 0.009&\ref{fig:abs3}\\

\noalign{\smallskip}
\hline

\end{tabular}\\
Notes: $^{*}$ Flare detected by eye,$^{a}$ full designation 1RXS J114728.8+664405,$^{b}$ full designation 2MASS J07471385+5020386, $^{+}$ the Pa$\beta$ line is not described well by one Gaussian.
\normalsize
\end{sidewaystable*}

\begin{sidewaystable*}
        \caption{\label{ew1} Gaussian fitting parameters for the H$\alpha$ and Pa$\beta$
        line for visually found Pa$\beta$ flares. }
\footnotesize
\begin{tabular}[h!]{lllccrrcrrcrrrrr}

\hline
\hline
\noalign{\smallskip}
	Karmn           & Name   & Spec.&JD  & $\lambda_ {\rm c}$  & Area & $\sigma$ &  $\lambda_ {\rm c}$   & Area & $\sigma$ &  $\lambda_ {\rm c}$   & Area & $\sigma$ & log($L_{\mathrm{\pab}}/L_{\mathrm{bol}}$) & $\Delta$pEW(\pab) &shown in  \\ 
                &     &type& -2450000&   &       &          & &   &   & \\
		&     & &[day] & [\AA] & [\AA] & [\AA] & [\AA] & [\AA] & [\AA] & [\AA] & [\AA] & [\AA] & & [\AA] & Fig.\\
\noalign{\smallskip}
\hline
\noalign{\smallskip}

	J01352$-$072 &Barta 161 12& M4.0\,V&7735.340&6565.77 & 4.41 & 1.64 & 6567.86 & 0.85 & 1.94 & 12823.60 & 0.18 & 2.14 & -6.12 & 0.009&\ref{fig:asym}\\
	J02088+494 &G 173-039 & M3.5\,V&7691.528 &  6564.99 & 1.75 & 0.64 & ...  & ...  & ... & 12822.50 & 0.08 & 1.32  & -5.20&0.009\\
	J02088+494 &G 173-039 & M3.5\,V&7987.588 & 6564.91 & 1.16 & 0.67 & 6565.25 & 0.94 & 1.81 & 12822.40 & 0.16 & 3.34 & -4.90&0.012\\
	J03473$-$019 &G 80-021  & M3.0\,V &7677.596 &  6564.61 & 0.92 & 0.57 & ... & ... & ... & 12821.70 & 0.06 & 0.57 & -5.54& 0.012\\
	J03473$-$019 &G 80-021  & M3.0\,V& 7766.314 &  6564.66 & 0.81 & 0.85 & ... & ... & ... & ... & ... &... & ...&0.008\\
	J07033+346 &LP 255-011& M4.0\,V& 8857.411 & 6564.61 & 1.90  & 0.61 & 6564.71 & 1.97 & 3.08 & 12821.60 & 0.15 & 1.40 & -6.22& 0.026\\
	J07472+503 &2MASS J0747+502$^{b}$& M4.0\,V&8882.442 & ... & ... & ...& ... & ... & ... & ...& ...& ... & ...&0.007\\ 
	J07558+833 &GJ 1101   & M4.5\,V&8041.580  & 6564.53 & 1.15 & 0.64 & 6564.42 & 1.19 & 2.96 & 12821.60 & 0.06 & 0.40 & -6.00 & 0.026& \ref{nobroad}\\ 
	J09161+018 &RX J0916.1+0153& M4.0\,V& 7712.658  & 6564.70 & 0.99 & 1.70 & 6565.64 & 1.04 & 5.36 &  12821.30 & 0.05 & 0.81 & -6.22 &0.031\\
	J11055+435 &WX UMa    & M5.5\,V&9748.384  &  6564.57 & 40.00 & 0.52 & 6565.54 & 29.87 & 1.62 & ... & ... & ... & ... &0.015\\
	J12156+526 &StKM2-809 & M4.0\,V& 7449.654 & ... & ... & ... & ... & ... & ... & ... & ... & ... & ...&0.005\\
	J12156+526 &StKM2-809 & M4.0\,V& 7558.430 &  ... & ... & ... & ... & ... & ... & ... & ... & ... & ...&0.023\\
	J12156+526 &StKM2-809 & M4.0\,V& 7752.752 & ... & ... & ... & ... & ... & ... & ... & ... & ... & ...&0.007\\
	J12428+418 &G 123-055 & M4.0\,V&7754.751 & 6564.60 & 2.05 & 0.58 & ...      & ...   & ...   &  12821.30 & 0.07 & 1.01 & -6.65 & 0.017&\ref{fig:notfound}\\
	J16570$-$043 &LP 686-027&M3.5\,V&7822.700 &  6564.74 & 1.72 & 0.63 & 6564.10 & 1.30 & 1.94 & 12821.60 & 0.10 & 1.07 & -7.15&0.018\\ 
	J22518+317 &GT Peg    &M3.0\,V&7762.276 &   6564.51 & 2.23 & 0.61 & 6565.89 & 1.14 & 1.64 & 12822.20 & 0.16 & 2.46 & -5.06&0.011\\ 
\noalign{\smallskip}
\hline

\end{tabular}\\
Notes: $^{*}$ Flare detected by eye,$^{a}$ full designation 1RXS J114728.8+664405,$^{b}$ full designation 2MASS J07471385+5020386
\normalsize
\end{sidewaystable*}

\end{document}